\documentclass[journal]{IEEEtran}
\usepackage[]{authblk}

\usepackage{graphicx,import}
\graphicspath{{img/}}
\usepackage{caption}
\usepackage{subcaption}
\usepackage{url}
\usepackage[numbers,sort&compress]{natbib}
\usepackage{booktabs}
\usepackage{array}
\usepackage{multirow}

\usepackage[inline]{enumitem}

\usepackage{tikz}
\usetikzlibrary{external}
\usepackage{pgfplots}
\usepgfplotslibrary{groupplots,dateplot}
\usetikzlibrary{patterns,shapes.arrows}
\pgfplotsset{compat=newest}
\usetikzlibrary{spy}
\usepgfplotslibrary{polar}
\usetikzlibrary{positioning}
\pgfplotsset{every axis plot/.append style={thick}}

\usepackage{float} 
\usepackage{placeins}

\usepackage{subcaption}
\usepackage{amsmath,amssymb,amsfonts,bm}
\usepackage{graphicx}
\usepackage{xcolor}
\usepackage{tabularx}
\usepackage[binary-units=true]{siunitx}
  \DeclareSIUnit{\belmilliwatt}{Bm}
\DeclareSIUnit{\dBm}{\deci\belmilliwatt}

\usepackage{listings}
\usepackage{comment}
\usepackage{import}
\usepackage{soul}

\usepackage{url}

\usepackage{orcidlink} 

\usepackage{hyperref} 

\usepackage[referable]{threeparttablex}
\usepackage{enumitem}
\usepackage[referable]{threeparttablex}
\renewlist{tablenotes}{enumerate}{1}
\makeatletter
\setlist[tablenotes]{label=\tnote{\alph*},ref=\alph*,itemsep=\z@,topsep=\z@skip,partopsep=\z@skip,parsep=\z@,itemindent=\z@,labelindent=\tabcolsep,labelsep=.2em,leftmargin=*,align=left,before={\footnotesize}}
\makeatother

\newcommand*{\vectr}[1]{\MakeLowercase{\boldsymbol{\mathrm{#1}}}}
\newcommand*{\matr}[1]{\MakeUppercase{\boldsymbol{\mathrm{#1}}}}
\newcommand*{\tran}[1]{{#1}^{\mkern-1.5mu\mathsf{T}}}
\newcommand*{\hermconj}[1]{{#1}^{H}}
\newcommand*{\inC}[1]{\in\mathbb{C}^{#1}}
\newcommand{\norm}[1]{\left\lVert#1\right\rVert}
\newcommand{\abs}[1]{\left\lvert#1\right\rvert}
\newcommand{\expt}[1]{\mathbb{E} \left\{#1\right\}}
\newcommand{\var}[1]{\mathbb{V} \left\{#1\right\}}

\newcommand{\normalized}[1]{\bar{#1}}

\newcommand{\Update}[2]{{\color{blue}#1 \st{#2}}}

\definecolor{colorULA}{rgb}{0.12157,0.46667,0.70588}
\definecolor{colorURA}{rgb}{1.00000,0.49804,0.05490}
\definecolor{coloriid}{rgb}{0.17255,0.62745,0.17255}
\definecolor{colorIID}{rgb}{0.17255,0.62745,0.17255}

\begin{document}

\title{Experimental Exploration of Unlicensed Sub-GHz Massive MIMO for Massive Internet-of-Things}

\author[*]{Gilles Callebaut\,\orcidlink{0000-0003-2413-986X}\thanks{Manuscript submitted to IEEE Transactions on Wireless Communications on May 26, 2021.}} 
\author[* $\dag$]{Sara Gunnarsson\,\orcidlink{0000-0002-5071-1631}} 
\author[*]{Andrea P. Guevara\,\orcidlink{0000-0002-7496-4772}}
\author[$\dag$]{\\Anders J Johansson\orcidlink{0000-0002-8351-4845}}
\author[* $\dag$]{Liesbet Van der Perre\,\orcidlink{0000-0002-9158-9628}} 
\author[$\dag$]{Fredrik Tufvesson\,\orcidlink{0000-0003-1072-0784}} 
\affil[*]{Department of Electrical Engineering, KU Leuven, Belgium}
\affil[$\dag$]{Department of Electrical and Information Technology, Lund University, Sweden}



\maketitle


\begin{abstract}
Due to the increase of Internet-of-Things (IoT) devices, IoT networks are getting overcrowded. Networks can be extended with more gateways, increasing the number of supported devices. However, as investigated in this work, massive MIMO has the potential to increase the number of simultaneous connections, while also lowering the energy expenditure of these devices.  We present a study of the channel characteristics of massive MIMO in the unlicensed sub-GHz band. The goal is to support IoT applications with strict requirements in terms of number of devices, power consumption, and reliability. The assessment is based on experimental measurements using both a uniform linear and a rectangular array. 
Our study demonstrates and validates the advantages of deploying massive MIMO gateways to serve IoT nodes. While the results are general, here we specifically focus on static nodes.
The array gain and channel hardening effect yield opportunities to lower the transmit power of IoT nodes while also increasing reliability. The exploration confirms that exploiting large arrays brings great opportunities to connect a massive number of IoT devices by separating the nodes in the spatial domain. In addition, we give an outlook on how static IoT nodes could be scheduled based on partial channel state information.


\end{abstract}\FloatBarrier







\begin{figure}[t]
     \centering
     
     \begin{subfigure}{0.9\linewidth}
         \centering
         \includegraphics[width=0.95\textwidth]{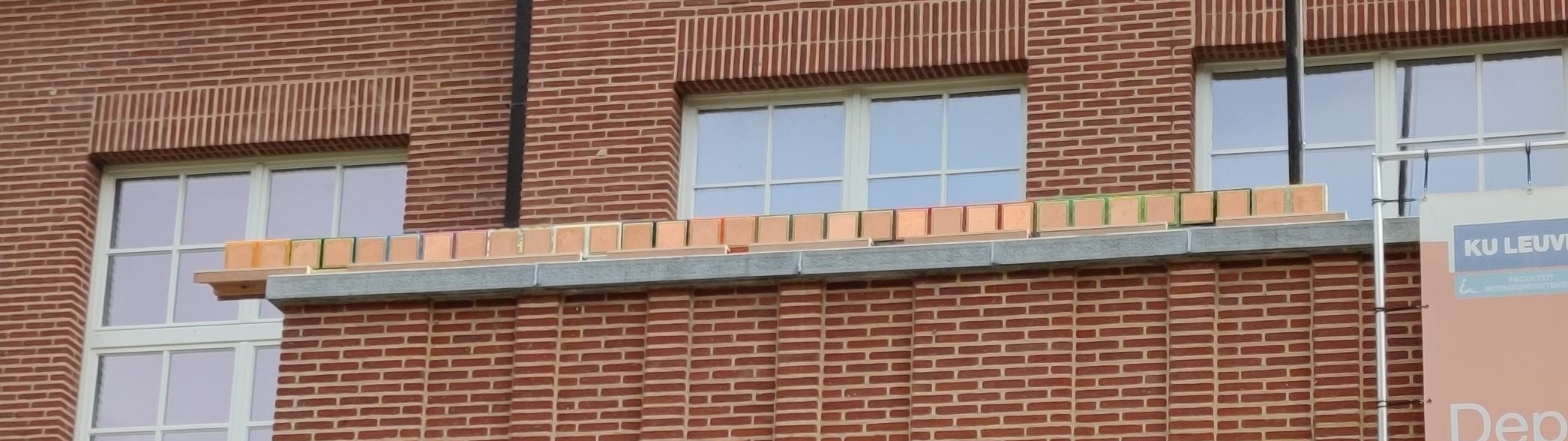}
         \caption{32-element ULA}\label{fig:ula}
     \end{subfigure}
     
     \begin{subfigure}{0.9\linewidth}
         \centering
         \includegraphics[width=0.95\textwidth]{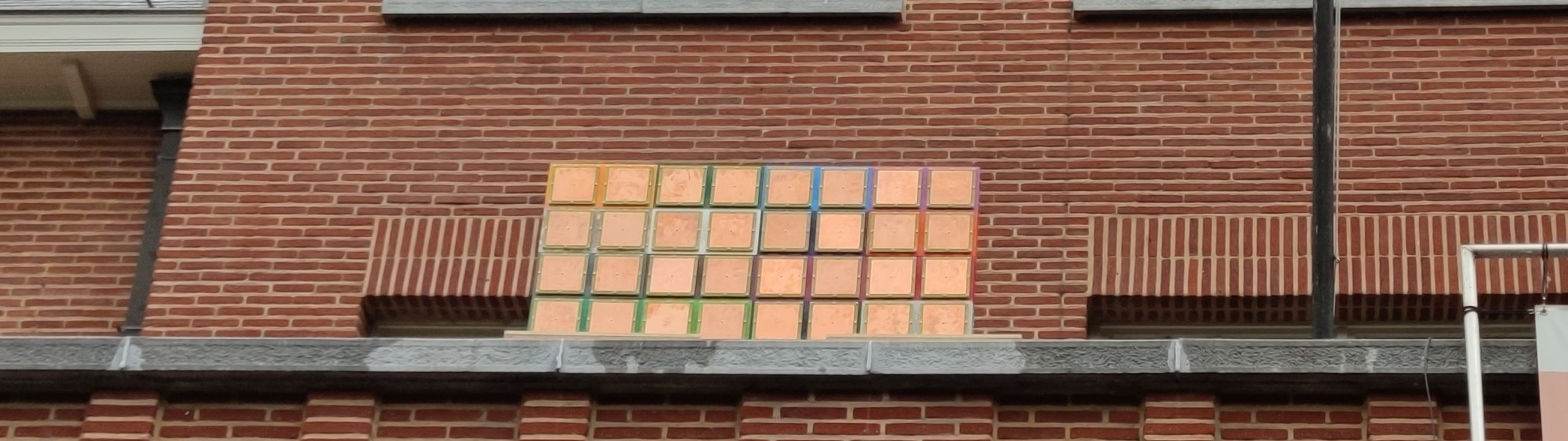}
         \caption{4 \(\times\) 8 URA}\label{fig:ura}
     \end{subfigure}
     \caption{Antenna array configurations.}\label{fig:conf}
\end{figure} 

\section{Towards massive and reliable IoT: the potential of multiple antenna systems}


\IEEEPARstart{I}{nternet-of-Things}
technology opens up a plethora of new applications and services in various domains. Examples include smart sustainable city services, precision farming, environmental monitoring and efficient utilities. 
These applications pose requirements on the wireless connectivity beyond what is offered by current networks. In particular, 
\begin{enumerate*}[label=(\roman*)]
    \item the projected massive number of devices to be supported,
    \item the stringent energy constraints of the IoT nodes, and
    \item the need to establish reliable connections,
\end{enumerate*}
ask for innovative wireless transmission approaches. Low-Power Wide-Area Networks (LPWANs) operating in unlicensed sub-GHz spectrum are of interest to many IoT applications.
They incur no or only a small subscription cost, and the operation at relatively low frequencies offers good coverage~\cite{cal2020}. 
In this paper we study the potential of deploying multiple antenna systems to upgrade these LPWANs to support future IoT services. We focus in particular on massive MIMO technology as it bears a great potential in view of the above listed requirements:
\begin{enumerate}
    \item It can support an unprecedented number of simultaneous connections through extensive spatial multiplexing.
    \item It typically operates with low complexity single antenna terminals. It allows to considerably reduce the transmit power at the node side thanks to the significant array gain at the base station side.
    \item It offers an increased reliability of the links due to the experienced channel hardening effect. 
\end{enumerate}


New approaches and technologies need to be adopted to accommodate the massive increase of IoT devices. In~\cite{matthieuAsilomar}, the authors use a maximum-likelihood strategy to decode two colliding users based on an interference model. This technique further extends the amount of supported devices in the network.
Other work has considered diversity techniques to improve IoT technologies. 
In~\cite{doi:10.1002/itl2.120} Snipe is introduced, which is an IoT system deploying two antennas to coherently combine the received signals at the gateway.  
Spatial diversity is exploited in~\cite{8480036} by coherently combining weak signals of different gateways in the cloud. The system, Charm, improves the range up to three times and the IoT battery-life fourfold. Including multiple antennas at the gateway is also proposed by~\cite{8372906} where this is theoretically analyzed. However, the authors consider only commercially available hardware and are therefore unable to use maximum ratio combining and are hence rather analyzing a selection combing technique.

The potential benefits of massive MIMO for IoT have been studied in theoretical work such as~\cite{7878690, BANA2019100859, 8241857, 8323218, 7842418, 8400514, senel2018grant}. Yet, whether this potential can be realized in real network deployments heavily relies on favorable characteristics of the radio propagation channels~\cite{6951994, doi:10.1002/9781119471509.w5GRef040}. In particular, many studies have assumed independent and identically distributed (i.i.d.) Rayleigh fading channels~\cite{BANA2019100859, senel2018grant, 8372906}. Measurement-based studies have shown that the reality deviates from this~\cite{9069181, 7062910, 7172496} and for example significant correlation is often observed over the antennas in the array~\cite{7062910}. Consequently, the channel hardening encountered is less pronounced than predicted by the  Rayleigh fading model~\cite{9069181}. Furthermore, different array topologies have been considered as they each favor different environments. 
For example, in \cite{8727203} it is demonstrated that an L-structured array outperforms both the ULA and the URA configuration. Hence, they advocate to also consider unconventional array structures.

Previous work has primarily focused on broadband transmission technologies~\cite{senel2018grant,9048663,beyene2015compressive,pereira2020large}. As a novel contribution, we study unlicensed narrowband low-power and long-range communication. Unlicensed massive MIMO has been studied in other work in the context of spectrum sharing~\cite{garcia2017massive}. However, to the best of our knowledge, massive MIMO to support LPWANs has not been considered. 

In this paper we report on the experimental campaign we have conducted to characterize narrowband sub-GHz channels with different configurations of a large antenna array. Based on the measured channel responses, we assess the potential of deploying massive MIMO in the unlicensed sub-GHz band for upgraded IoT connectivity. The system architecture and implementation is further elaborated on in~\cite{asilomar}.

To summarize, our contributions are (i) channel propagation study in sub-GHz for massive MIMO and (ii) study of massive MIMO for LPWANs. We demonstrate -- based on the measurements -- that the reliability and the number of simultaneous connections is increased. Furthermore, the array gain can extend the coverage of the base station or allows to reduce the transmit power of the IoT devices.

This paper is further organized as follows. In the next section we introduce the system model and theoretic fundamentals. Section~\ref{sec:setup} introduces the measurement setup and scenarios. In Section~\ref{sec:exploration}, we present the exploration and assessment performed based on the experiments.
Finally, in Section~\ref{sec:concl} the main conclusions of this paper are summarized and an outlook on future progress towards massive MIMO-upgraded networks for future IoT is given.

While commonly the single-antenna device is denoted as the User Equipment (UE) in a massive MIMO or LTE context, in an Internet-of-Things setting the term \textit{node} or \textit{IoT node} is used. As we study the effects of massive MIMO in an Internet-of-Things setting, we will use the node to depict the end-device or single-antenna device.

\section{System Model and Theoretic Fundamentals}

\begin{table}[tbp]
\centering
\caption{Measurement Setup}
\label{tab:measurement-setup}
\centering
  \begin{threeparttable}
\begin{tabular}{@{}lll@{}}
\toprule
Parameter &  Symbol & Value\\ \midrule
Carrier frequency & \(f_c\) & \SI{869.525}{\mega\hertz} \\
Number of subcarriers (\SI{15}{\kilo\hertz}) & \(F\) & 2 \\
Number of snapshots & \(N\) & 1000\tnotex{tnote:static} /6000\tnotex{tnote:continuous} \\
Transmit power (coerced) &\(P_{tx}\)& \SI{22.6}{\dBm}\\
Number of bBase station antennas &\(M\)& 32 \\
Number of IoT nodes &\(K\)& - \\
Base station array configuration && ULA/URA \\
Type of BS antenna && Patch \\
Type of UE antenna && Dipole \\
Sample interval && \SI{10}{\milli\second}\\
Sample duration && \SI{66.67}{\micro\second}\\
Modulation & & OFDM\\
Subcarrier modulation && QPSK \\ 
UE height && \SI{1.5}{\meter} \\
BS height && \SI{7}{\meter} \\
Antenna polarization && vertical  \\\bottomrule
\end{tabular}
\begin{tablenotes}
      \item\label{tnote:static} For static measurements.
      \item\label{tnote:continuous} For continuous measurements.
    \end{tablenotes}
\end{threeparttable}
\end{table}

Throughout the paper, vectors are denoted by boldface lower case (\(\vectr{x}\)) and matrices by boldface capital letters~(\(\matr{X}\)).
The superscripts \(\tran{(\cdot)}\) and \(\hermconj{(\cdot)}\) are used to denote the transpose and the conjugate transpose operations. 
The absolute value is denoted by \(\abs{\cdot} \), \(\norm{\cdot}\)  denotes the \(\ell_2\) norm and \(\norm{\cdot}_F\) is the Frobenius norm or Euclidian norm.
The notations \(\expt{\cdot}\) and \(\var{\cdot}\) denote the expectation and the variance of a random variable, respectively. An overline, e.g., \(\normalized{\vectr{x}}\), indicates a normalized quantity. 
The eigenvalues of a matrix are obtained by the operator \(\lambda \left(\cdot\right) \). The optional subscripts \(\min\) and \textit{max} are used to get the minimum and maximum value. The set of complex numbers is denoted by the symbol \(\mathbb{C}\). 

\textbf{Channels.} The channels are estimated for each base station (BS) antenna at different positions and for different frequency points and time instances. An overview of the system parameters and used symbols can be found in Table~\ref{tab:measurement-setup}. The total number of antennas is denoted by \(M\). The subscript \(m\) specifies the antenna index. Similarly, the total number of node positions is denoted by \(K\) and a given position index by \(k\). As elaborated in~\cite{asilomar}, the channel is estimated over two frequency points \(F\) at different time instances \(N\), with \(f\) and \(n\) denoting the frequency point and time instance index, respectively. Consequently, a channel matrix for a position \(k\) is expressed as \(\matr{H}_{k} \inC{N \times F \times M}\).
Note that the collected channel includes both small-scale and large-scale fading as well as potential effects from the hardware and interfering devices.
As a baseline to compare our results, we use the i.i.d. complex Gaussian channel, i.e., Rayleigh fading channel. This is modeled as a complex random variable with zero mean and unit variance of power (\(\mathit{h}\sim\mathcal{C}\mathcal{N}\left(0,1\right)\)).

\textbf{Channel Estimation.}
The channel is estimated by sending a unique pilot signal from each single-antenna device \(d\) to the base station.
Here, each device sends a dedicated pilot signal with a dedicated frequency, i.e., during the pilot phase all devices use distinct frequencies. The base station can estimate the channel response for each device \(d\) at a specific location \(k\) based on the \textit{a~priori} known pilot sequence. The channel is derived by correlating the complex conjugate of each pilot with the received uplink signal at their respective carrier frequency:
\begin{equation}
\hat{\vectr{h}}_{m,k} = \vectr{y}_m \hermconj{\vectr{\phi}_{m,d}} \enspace.
\end{equation}
We removed the dependency on \(d\) from the channel response as only one device was used during the measurements.  For further analysis and readability, we will be using \(\vectr{h}\) to denote the estimated noisy channel \(\hat{\vectr{h}}\).

\textbf{Normalization.}
The channel \(h_k(n,f)\) at each position \(k\) is normalized such that the average channel gain over all antennas, frequencies and snapshots is equal to one, i.e., \(\norm{\normalized{h}_{k}} = 1\): 

\begin{equation}\label{eq:norm}
    \normalized{\vectr{h}}_{k}(n,f) =  \frac{\vectr{h}_k(n,f)}{\sqrt{\frac{1}{N F M} \sum_{n=1}^{N} \sum_{f=1}^{F} \sum_{m=1}^{M} \abs{h_{k,m}\left(n,f\right)}^2}}.
\end{equation}





\textbf{Channel hardening.} We assess the channel hardening as a representative characteristics for the decrease of the fading with increasing number of antennas, by which the channel becomes more deterministic and the reliability of the link improves. According to \cite{downlink_pilots}, a channel $\mathbf{\normalized{h}}_{k}$ offers channel hardening if

\begin{equation}
\frac{\var{\|\mathbf{\normalized{h}}_{k}\|^2}}{\expt{\|\mathbf{\normalized{h}}_{k}\|^2}^2}\rightarrow 0, \hspace{0.5cm} \text{as} \hspace{0.2cm} M\rightarrow \infty,
\label{eq:chhard_def}
\end{equation}

where the variance and expectation is taken over the frequency and time dimensions for a given position \(k\). This means that as the number of antennas increases, the variation of channel gain decreases. Here, the standard deviation is considered, as is also done in \cite{9069181}.  
This means that, for a subset of $M$ base station antennas and each user, the instantaneous channel gain, at a given time \(n\) and frequency \(f\), is defined as

\begin{equation}
\overline{G}_k(n,f) = \frac{1}{M}\sum_{m=1}^{M} |\overline{h}_{k,m}(n,f)|^2,
\label{eq:subset_pwr}
\end{equation}

\noindent resulting in an average channel gain of

\begin{equation}
\mu_k = \frac{1}{N F} \sum_{n=1}^N \sum_{f=1}^F \overline{G}_k(n,f) = 1, 
\label{eq:mean}
\end{equation}

\noindent which no longer depends on the number of antennas at the base station as we averaged with respect to all antennas \(M\). Note, a distinction is made between the channel gain and the array gain. The channel gain is the absolute value of the channel coefficient squared. In contrast, an array gain is, here, the sum of the channel gains of each antenna relative to the channel gain of a single antenna case. The array gain implies an additional gain of having multiple antennas, i.e., an array.

Finally, the standard deviation of the channel gain at a given position \(k\) can be computed as

\begin{equation}
\sigma_k = \sqrt{\frac{1}{N F} \sum_{n=1}^{N} \sum_{f=1}^{F} |\overline{G}_k(n,f) - \mu_k|^2},
\label{eq:std}
\end{equation}

\noindent which is used to quantify the channel hardening by taking the difference between the complete set of base station antennas and one base station antenna.


\textbf{Correlation coefficient.}
To compare the channel correlation between two channels, the correlation coefficient\footnote{To be complete, the correlation coefficient here defined is not the same as the correlation coefficient usually used in statistics.} is studied. It is defined between two channel vectors \(\vectr{h}_{i}\) and \(\vectr{h}_{j}\) as
\begin{equation}\label{eq:corr-coeff}
    \delta_{i,j}(n,f) = 
    \frac
    {\abs{
    \hermconj{\normalized{\vectr{h}}_{i}(n,f)} \cdot \normalized{\vectr{h}}_{j}(n,f)
    }}
    {\norm{\normalized{\vectr{h}}_{i}(n,f)} \norm{\normalized{\vectr{h}}_{j}(n,f)}},
\end{equation}
where the channels are normalized according to~(\ref{eq:norm}). The correlation coefficient is estimated by picking two random measurement locations (\(i\) and \(j\)). From these locations, we select one random snapshot (\(n\)) and frequency point (\(f\)) and use these channel vectors to compute the correlation coefficient. The correlation coefficient depicts the antenna-averaged channel correlation between two channel instances. In case of uncorrelated channels, the coefficient is zero, while the coefficient is one for channels which are parallel, i.e., equal up to a scaling factor.

\textbf{Channel correlation.}
The channel correlation matrix per position is obtained as the mean over \textit{N} snapshots of the user channel and its channel conjugate transpose for the same number of samples.
\begin{equation}\label{eq.R}
    \matr{R} = \frac{1}{N} \sum_{n = 1}^{N} {\normalized{\vectr{h}}(n,f) \hermconj{\normalized{\vectr{h}}}(n,f)},
\end{equation}

where \(\matr{R}\inC{M\times  M}\) and  \(\vectr{h}(n) \inC{M \times N}\). It expresses the correlation between the channels observed by each antenna. 

\textbf{The condition number.}
The joint orthogonality of multiple positions, or channels, is investigated by using the condition number~\cite{4217646,6328480}. 
It is defined as
\begin{equation}
    \kappa_{K,M}(n,f) = \frac{\lambda_{\text{max}}(\hermconj{\normalized{\matr{H}}(n,f)} \normalized{\matr{H}}(n,f))}{\lambda_{\text{min}}(\hermconj{\normalized{\matr{H}}(n,f)} \normalized{\matr{H}}(n,f))},
\end{equation}
 with \(\kappa_{K,M} \in [1, \infty)\). The channel matrix \(\normalized{\matr{H}}(n,f)\) consists of the normalized channel instances of \(K\) positions \([\normalized{\vectr{h}}_{1}(n,f) \dots \normalized{\vectr{h}}_{K}(n,f)] \inC{M \times K}\). 
 A high condition number means
that at least one pair of channels is strongly correlated. When all channels are pairwise orthogonal the condition number becomes one. The inverse condition number is chosen as a metric to be able to express the orthogonality in a finite range, i.e., \(\kappa_{K,M}^{-1} \in [0,1]\). 

\textbf{Chordal distance.}
The chordal distance measures the orthogonality between two eigenspaces, represented as the \textit{p-}dominant eigenvectors. By definition~\cite{massivemimobook}, the chordal distance between two matrices $\matr{U}_i$ and $\matr{U}_j$ is given by 
\begin{equation}\label{eq.chordal}
    d_c\left(\matr{U}_i,\matr{U}_j\right) = \norm{\matr{U}_i \hermconj{\matr{U}}_i - \matr{U}_j\hermconj{\matr{U}}_j}^2 _F ,
\end{equation}
where $\matr{U}_k \inC{M\times \mathit{p}}$ and $\matr{U}_i \inC{M\times \mathit{p}}$ are the unitary matrices that span over the number of antennas \textit{M} and  \textit{p}-dominant eigendirections. The unitary matrices are obtained in the eigenvalue decomposition from the Hermitian matrix $\matr{R}$ as follows: $\matr{R} = \matr{U}\matr{D}\hermconj{\matr{U}}$.
Note that in order to estimate the chordal distance it is important to  obtain the \textit{p}-dominant eigendirections in advance.

\begin{figure}[tbp]
    \centering
    \includegraphics[width=0.9\linewidth]{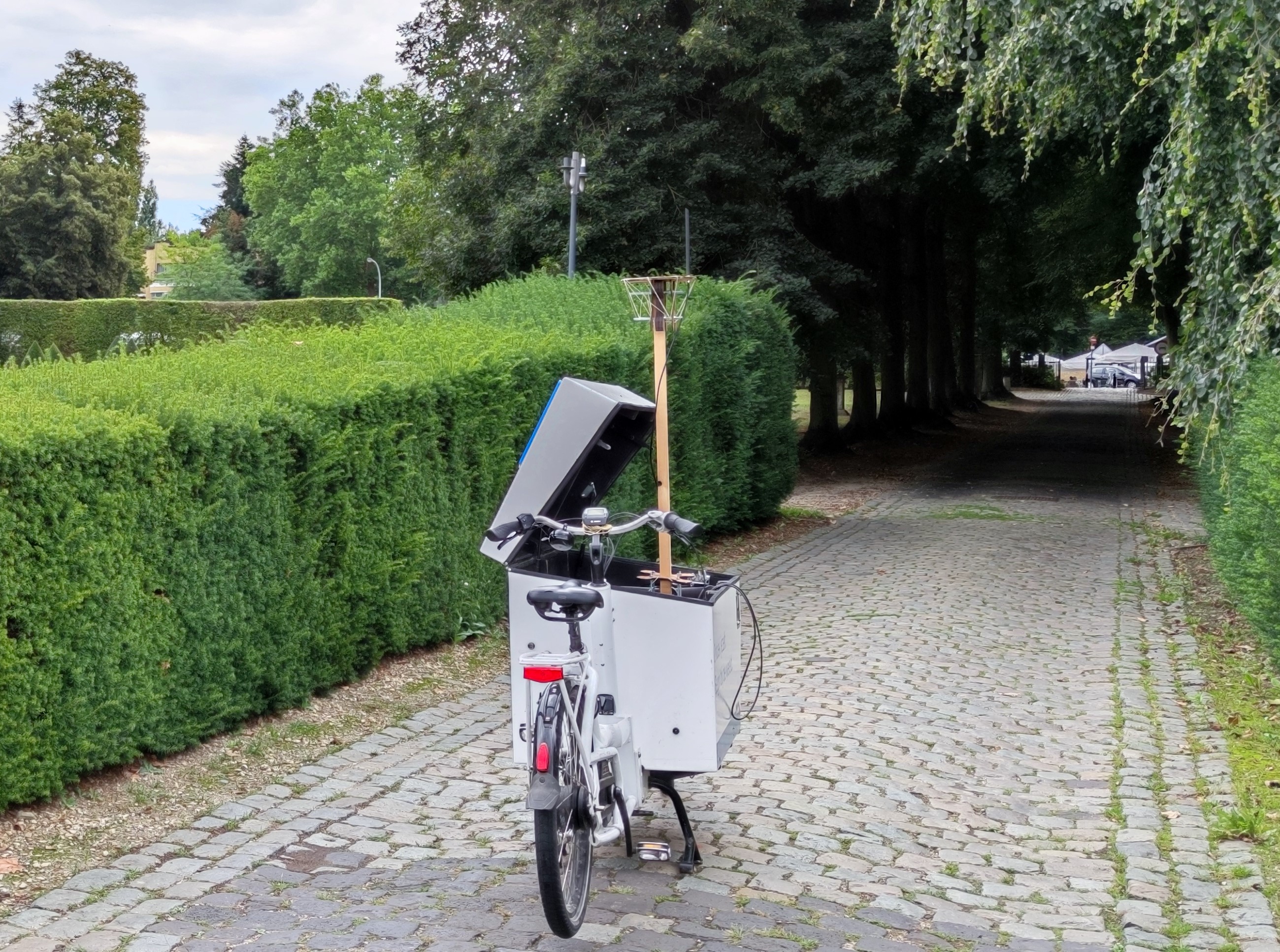}
    \caption{Transmit node with a single dipole antenna.}
    \label{fig:user}
\end{figure}



\section{Measurement Setup and Scenario}\label{sec:setup}

The 5G massive MIMO testbed at KU Leuven\footnote{\url{https://www.esat.kuleuven.be/telemic/research/NetworkedSystems/infrastructure/massive-mimo-5g}} based on National Instruments equipment was used during these experiments. The testbed runs the LabVIEW Communications MIMO Application. This application was designed for LTE-like transmission. Therefore, the LabView application had to be adapted to conform to the regulations of~\cite{cept2017erc}. To be precise, the occupied bandwidth, the transmit power and duty cycle of the framework had to be altered. This is elaborated on in~\cite{asilomar}.
Table~\ref{tab:measurement-setup} summarizes the measurement setup. 

The base station is, as mentioned before, equipped with \num{32} vertically polarized patch antennas.
Two array configurations were used, a 4-by-8 URA (Fig.~\ref{fig:ura}) and a 32-element ULA (Fig.~\ref{fig:ula}). The patch antenna was designed to operate in the \SI{868}{\mega\hertz} band. In theoretical massive MIMO papers a 32-element array may be considered relatively small, yet in absolute terms at this operation frequency the array is definitely quite large, i.e., \SI{5.5}{\meter} for a 32-element ULA at \SI{868}{\mega\hertz} or with a wavelength of \SI{34.5}{\centi\meter}.
The array consists of two-element holders. These holders facilitate the design of different array topologies. While in this measurement we use a rectangular and a linear uniformal array, cylindrical and distributed arrays can be easily constructed with this design. Furthermore, the holders ensure that all antennas are half-wavelength spaced when connected.
At the node side, a single dipole antenna is used (Fig.~\ref{fig:user}).

\begin{figure}[tbp]\centering
\fontsize{6pt}{10pt}\selectfont
    \def\svgwidth{0.9\columnwidth}
\begingroup%
  \makeatletter%
  \providecommand\color[2][]{%
    \errmessage{(Inkscape) Color is used for the text in Inkscape, but the package 'color.sty' is not loaded}%
    \renewcommand\color[2][]{}%
  }%
  \providecommand\transparent[1]{%
    \errmessage{(Inkscape) Transparency is used (non-zero) for the text in Inkscape, but the package 'transparent.sty' is not loaded}%
    \renewcommand\transparent[1]{}%
  }%
  \providecommand\rotatebox[2]{#2}%
  \newcommand*\fsize{\dimexpr\f@size pt\relax}%
  \newcommand*\lineheight[1]{\fontsize{\fsize}{#1\fsize}\selectfont}%
  \ifx\svgwidth\undefined%
    \setlength{\unitlength}{971.9502088bp}%
    \ifx\svgscale\undefined%
      \relax%
    \else%
      \setlength{\unitlength}{\unitlength * \real{\svgscale}}%
    \fi%
  \else%
    \setlength{\unitlength}{\svgwidth}%
  \fi%
  \global\let\svgwidth\undefined%
  \global\let\svgscale\undefined%
  \makeatother%
  \begin{picture}(1,0.83844001)%
    \lineheight{1}%
    \setlength\tabcolsep{0pt}%
    \put(0,0){\includegraphics[width=\unitlength,page=1]{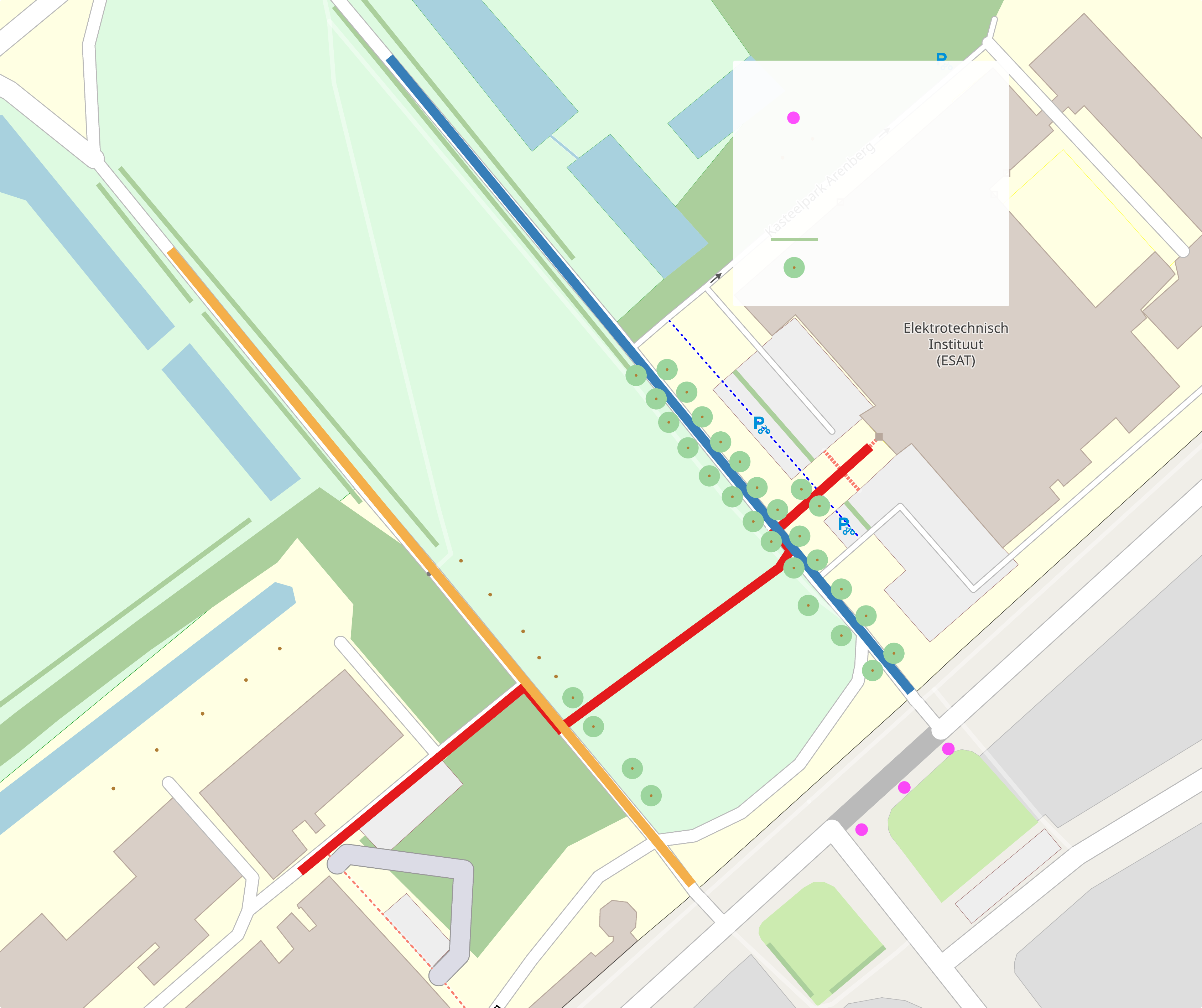}}%
    \put(0.69961805,0.7590149){\makebox(0,0)[lt]{\lineheight{1.25}\smash{\begin{tabular}[t]{l}Base Station\end{tabular}}}}%
    \put(0.69961805,0.73451164){\makebox(0,0)[lt]{\lineheight{1.25}\smash{\begin{tabular}[t]{l}Points D\end{tabular}}}}%
    \put(0.69961805,0.71000841){\makebox(0,0)[lt]{\lineheight{1.25}\smash{\begin{tabular}[t]{l}Path A\end{tabular}}}}%
    \put(0.69961805,0.68550512){\makebox(0,0)[lt]{\lineheight{1.25}\smash{\begin{tabular}[t]{l}Path B\end{tabular}}}}%
    \put(0.69961805,0.66100189){\makebox(0,0)[lt]{\lineheight{1.25}\smash{\begin{tabular}[t]{l}Path C\end{tabular}}}}%
    \put(0,0){\includegraphics[width=\unitlength,page=2]{img/map_campus.pdf}}%
    \put(0.69851352,0.63524049){\makebox(0,0)[lt]{\lineheight{1.25}\smash{\begin{tabular}[t]{l}Hedge\end{tabular}}}}%
    \put(0.69792825,0.61000076){\makebox(0,0)[lt]{\lineheight{1.25}\smash{\begin{tabular}[t]{l}Tree\end{tabular}}}}%
    \put(0,0){\includegraphics[width=\unitlength,page=3]{img/map_campus.pdf}}%
  \end{picture}%
\endgroup%
\caption{Overview of rural measurement area. All paths have a length of approximately \SI{140}{\meter}.}\label{fig:map}
\end{figure}

The measurements were conducted in front of the Department of Electrical Engineering (ESAT) building in Heverlee, Belgium. The measurement environment can be seen in Fig.~\ref{fig:map}.
The base station was placed on the balcony at the first floor of the building at a height of \SI {7}{\metre}.

During the experimental campaign, we collected the channels on all \num{32} antennas\footnote{Due to an unforeseen software issue, only the first \num{31} antennas were usable.} for static and continuous measurements. 
Between each static measurement position we moved the transmit antenna (node) \num{10} meters according to the paths shown in Fig.~\ref{fig:map}. The paths were chosen in order to have positions perpendicular and parallel to the base station, as well as having NLoS and LoS positions. As the measurements were done in summer, the presence of foliage on the trees has a non-negligible effect on the collected channels. Both the measurement data\footnote{\url{dramco.be/massive-mimo/measurement-selector/\#Sub-GHz}} and the processing scripts\footnote{\url{github.com/GillesC/MARRMOT}} are available in open-source.

\begin{figure}[tbp]
    \centering
    \includegraphics[width=0.8\linewidth]{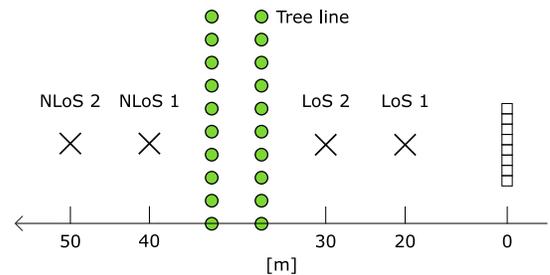}
    \caption{Points considered to be LoS and NLoS including distances (in \si{\meter}) with respect to the base station.}
    \label{fig:los-nlos-points}
\end{figure}

When comparing LoS and NLoS scenarios, we use two static points per case. 
The locations of the measurement points are shown in Fig.~\ref{fig:los-nlos-points} and further used in Section~\ref{sec:exploration} to study the effect of LoS and NLoS scenarios.






\section{Evaluation}\label{sec:exploration}


We here evaluate different performance metrics, grouped based on their impact on crucial aspects for IoT communication. Section~\ref{sec:reliability-coverage-EE} explores the benefits of massive MIMO regarding the reliability, coverage and energy efficiency of the IoT nodes. The ability to serve multiple nodes and a first look on how to schedule them is studied in Sections~\ref{sec:serve-nodes} and~\ref{sec:schedule-nodes}.

When investigating the impact of the number of base station antennas, we selected \(M\) subsequent array elements according to their numbering. In the case of a ULA this always results in a ULA of \(M\) antenna elements, this in contrast to selecting random antenna elements. The same reasoning does not always hold for the URA configuration. An example where this does not hold is shown in Fig.~\ref{fig:antenna-numbering}. Consequently, the results of the URA configuration, with respect to the number of antennas, needs to be carefully interpreted. 
To be able to capture small-scale fading, while still have a negligible effect of the large-scale fading, we divided the continuous measurements in paths with a length of approximately \(25\, \lambda\).






\subsection{Increased Energy Efficiency, Coverage and Reliability}\label{sec:reliability-coverage-EE}
The reliability, coverage and energy efficiency improvement is studied by means of the channel gain diversity, combining gain and channel hardening effect.
The combining gain allows to reduce the transmit power of the nodes due to the increased gain when combining the many base station antennas; Or equivalently, the coverage can be extended by the achieved combined gain.
Further on, when a channel offers channel hardening, the variance of the channel gain decreases as the number of antennas increases; hence, providing a more reliable channel. As a result, the fading margin at the IoT node can also be reduced. As the channel becomes more deterministic, the probability of packet losses decreases, and thereby the number of required retransmissions is reduced. 

\begin{figure*}[tbp]
    \centering
    \includegraphics[width=0.9\textwidth]{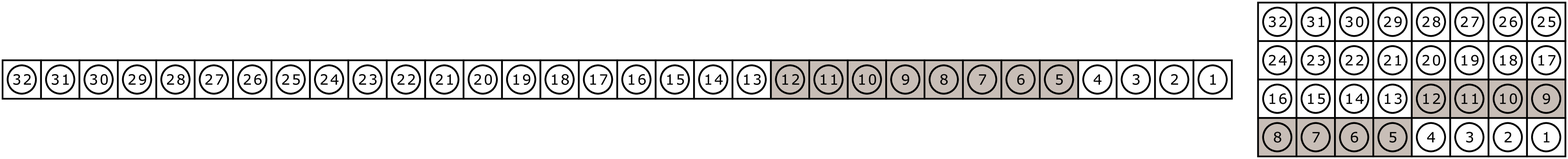}
    \caption{Antenna numbering with \num{8} randomly selected subsequent antenna elements (highlighted).}
    \label{fig:antenna-numbering}
\end{figure*}

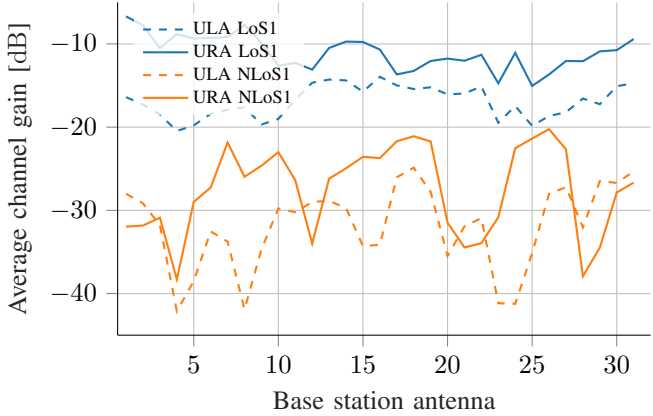
\begin{figure}[h]
    \centering
%
%
\definecolor{colorLoS}{rgb}{0.12157,0.46667,0.70588}
\definecolor{colorNLoS}{rgb}{1.00000,0.49804,0.05490}

\begin{tikzpicture}

\begin{axis}[%
axis lines* = {left},
legend cell align={left},
legend style={fill opacity=0.8, draw opacity=1, text opacity=1, at={(0.03000,0.97000)}, anchor=north west, draw=white!80.00000!black},
legend style={nodes={scale=0.7, transform shape}},
legend style={draw=none},
tick pos=left,
width = 0.8\linewidth,
height = 0.5\linewidth,
scale only axis,
xmin=0.5,
xmax=32,
xlabel style={font=\color{white!15!black}},
xlabel={Base station antenna},
ymin=-45,
ymax=-5,
ylabel style={font=\color{white!15!black}},
ylabel={Average channel gain [\si{\deci\bel}]},
tick align=outside,
axis background/.style={fill=white},
axis x line*=bottom,
axis y line*=left,
xmajorgrids,
ymajorgrids,
]
\addplot [color=colorLoS, dashed]
  table[row sep=crcr]{%
1	-16.3972377942314\\
2	-17.2856147019941\\
3	-18.47088989478\\
4	-20.468972283381\\
5	-19.7868148053648\\
6	-18.4061274116789\\
7	-17.9027925115464\\
8	-17.6112492515215\\
9	-19.6828536637749\\
10	-19.0103395705234\\
11	-16.6942323487212\\
12	-14.6596431024805\\
13	-14.2820521895387\\
14	-14.3822899674349\\
15	-15.7336534840641\\
16	-13.9438678175516\\
17	-14.9833269171677\\
18	-15.4204499123627\\
19	-15.2104637289196\\
20	-16.0638159427955\\
21	-15.9543479399569\\
22	-15.1103008318082\\
23	-19.4687771069365\\
24	-17.4650258589548\\
25	-19.8417926949612\\
26	-18.6614785478897\\
27	-18.2034972589624\\
28	-16.563777082697\\
29	-17.2394002757141\\
30	-15.0893684056684\\
31	-14.7030791061941\\
};
\addlegendentry{ULA LoS1}

\addplot [color=colorLoS]
  table[row sep=crcr]{%
1	-6.69352535181316\\
2	-7.76871051737998\\
3	-10.5141369436943\\
4	-8.81069175536649\\
5	-9.34539499516342\\
6	-9.27693807762921\\
7	-9.17180913951639\\
8	-7.72307574155668\\
9	-9.87515228030174\\
10	-12.6198896784583\\
11	-12.2708968787035\\
12	-13.0876676853786\\
13	-10.4819123587122\\
14	-9.71918030668194\\
15	-9.76376223975364\\
16	-10.67337499801\\
17	-13.6667984554291\\
18	-13.2430001745028\\
19	-12.0457355475536\\
20	-11.7617471746585\\
21	-12.0188151339036\\
22	-11.3046785936492\\
23	-14.7304988883926\\
24	-11.0678573884583\\
25	-15.0500646541943\\
26	-13.667232336345\\
27	-12.0499912067438\\
28	-12.0747416136724\\
29	-10.8950989743878\\
30	-10.7473031658184\\
31	-9.41606269181901\\
};
\addlegendentry{URA LoS1}

\addplot [color=colorNLoS, dashed]
  table[row sep=crcr]{%
1	-28.0083128580877\\
2	-29.1695730688361\\
3	-31.7831184554716\\
4	-42.1159700764022\\
5	-38.5049394636728\\
6	-32.5278162205052\\
7	-33.7501371413626\\
8	-41.902664768214\\
9	-34.6818556889079\\
10	-29.7744413718188\\
11	-30.2206372104263\\
12	-28.9860281056434\\
13	-28.8646261325471\\
14	-29.7193397582438\\
15	-34.2958429194143\\
16	-34.1396661706205\\
17	-26.0250013196766\\
18	-24.8673411640888\\
19	-27.7720609037921\\
20	-35.470781989245\\
21	-31.9200808641992\\
22	-30.9290662716493\\
23	-41.1628634932813\\
24	-41.2374443390839\\
25	-35.128603160169\\
26	-28.0117178115684\\
27	-27.2164637426539\\
28	-32.0403707403844\\
29	-26.4771215803383\\
30	-26.7024059645954\\
31	-25.3219759405751\\
};
\addlegendentry{ULA NLoS1}

\addplot [color=colorNLoS]
  table[row sep=crcr]{%
1	-31.9582175464254\\
2	-31.8240291008751\\
3	-30.8822412458664\\
4	-38.2930366726976\\
5	-28.9780178337589\\
6	-27.2586609974074\\
7	-21.8397749468845\\
8	-25.9636103547777\\
9	-24.6385824457947\\
10	-23.0032632879129\\
11	-26.3841924921691\\
12	-33.9948299908295\\
13	-26.1936576111183\\
14	-24.945196980159\\
15	-23.5618589223869\\
16	-23.7238430759187\\
17	-21.6920782304119\\
18	-21.0998061766624\\
19	-21.7193612762789\\
20	-31.4848418313587\\
21	-34.4590948063332\\
22	-33.9468925373877\\
23	-30.8334312175626\\
24	-22.5344767483957\\
25	-21.370125483023\\
26	-20.2466402820434\\
27	-22.6687950721618\\
28	-37.9183251967322\\
29	-34.4800630781099\\
30	-27.8551392729634\\
31	-26.6628181669223\\
};
\addlegendentry{URA NLoS1}

\end{axis}
\end{tikzpicture}%
    \caption{The average (over time) channel gain per antenna element demonstrates the presence of large-scale fading over the antennas. The antenna numbering is according to Fig.~\ref{fig:antenna-numbering}.}%
    \label{fig:large-scale}
\end{figure}

\begin{figure}[h]
    \centering
\begin{tikzpicture}

\begin{axis}[
axis lines* = {left},
colorbar,
colorbar style={align=center, title={\footnotesize Average channel\\\footnotesize gain (\si{\deci\bel})}},
colormap={mymap}{[1pt]
  rgb(0pt)=(0.61961,0.00392,0.25882);
  rgb(1pt)=(0.83529,0.24314,0.30980);
  rgb(2pt)=(0.95686,0.42745,0.26275);
  rgb(3pt)=(0.99216,0.68235,0.38039);
  rgb(4pt)=(0.99608,0.87843,0.54510);
  rgb(5pt)=(1.00000,1.00000,0.74902);
  rgb(6pt)=(0.90196,0.96078,0.59608);
  rgb(7pt)=(0.67059,0.86667,0.64314);
  rgb(8pt)=(0.40000,0.76078,0.64706);
  rgb(9pt)=(0.19608,0.53333,0.74118);
  rgb(10pt)=(0.36863,0.30980,0.63529)
},
legend style={nodes={scale=0.7, transform shape}},
point meta max=-6.82519,
point meta min=-44.90955,
tick align=outside,
tick pos=left,
width = .8\linewidth,
x dir=reverse,
x grid style={white!69.01961!black},
xmin=-0.50000, xmax=30.50000,
xtick style={color=black},
ymin=-0.50000, ymax=47.50000,
axis line style={draw=none},
tick style={draw=none},
yticklabels={,,},
xticklabels={,,},
ylabel={Measurement point},
xlabel={Base station antenna}
]
\addplot graphics [includegraphics cmd=\pgfimage,xmin=-0.50000, xmax=30.50000, ymin=47.50000, ymax=-0.50000] {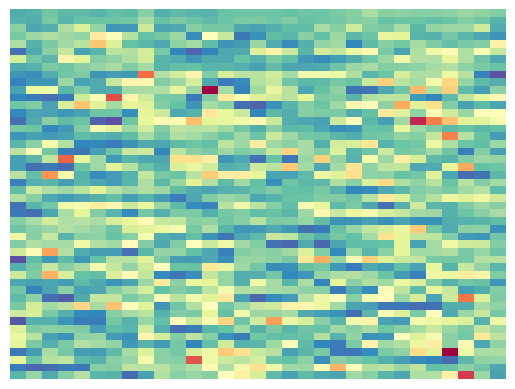};
\end{axis}

\end{tikzpicture}
    \caption{Average channel gain per base station antenna (ULA). Each row depicts the average channel gain per measurement point for each antenna element.}%
    \label{fig:waterfall}
\end{figure}

\textbf{Channel Gain Diversity over the Array.}
In \cite{7062910}, it was observed that antenna arrays with a large aperture experience antenna-dependent large-scale fading. In this work, we have observed similar behavior even with a low number of antennas. Notably, the physical aperture of the array is larger for \SI{868}{\mega\hertz} than \SI{2.6}{\giga\hertz} as in~\cite{7062910}. The un-normalized average channel gain, summed over frequency and per base station antenna, is depicted in Fig.~\ref{fig:large-scale} for the two array configurations and a LoS and NLoS scenario, respectively.
By averaging the channel gain over time for each antenna element, we take away the small-scale fading present on each antenna. 
Fig.~\ref{fig:large-scale} shows that the large-scale can not be considered constant over the antennas as is frequently assumed in theoretical work~\cite{senel2018grant}. Depending on the position of the antenna in the array, it can be shadowed or see different multi-path components; this naturally becomes even more noticeable in the NLoS scenario for both array configurations. 
Moreover, the average observed difference between the maximum and minimum channel coefficients between two antennas 
during one measurement is \SI{15.4}{\deci\bel} and \SI{13.2}{\deci\bel} for the ULA and URA, respectively.
Depending on the location of the node and how the multi-path components travel in the environment, the  antennas at the base station will hence not contribute equally to the overall received signal. Fig.~\ref{fig:waterfall} illustrates this by showing the average channel gain for all base station antennas and all measured points. The considerable differences demonstrate that spatial diversity can significantly improve the link reliability. Hereby it can also be noted that at some points, a specific antenna can be in a fading dip while  being one of the strongest antennas at other points. 

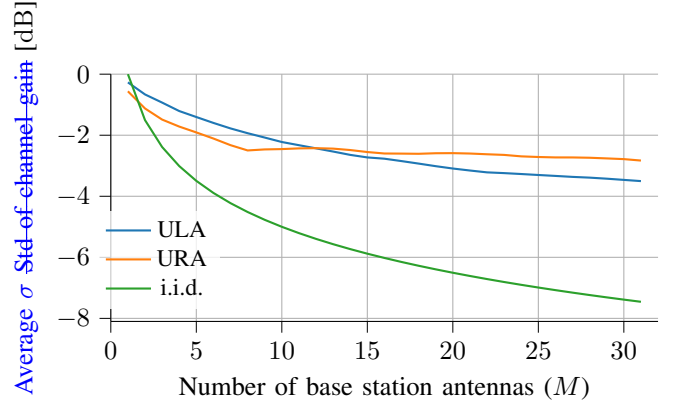
\begin{figure}[tbp]
    \centering
    \begin{tikzpicture}

\begin{axis}[%
xmin=0,
xmax=32,
ymin=-8.1,
ymax=0,
ylabel={\Update{Average \(\sigma\)}{Std of channel gain} [\si{\deci\bel}]},
axis lines* = {left},
legend style={fill opacity=0.8, draw opacity=1, text opacity=1, draw=white!80.00000!black, at={(0.2,0.45)}},
legend style={nodes={scale=0.9, transform shape}},
legend style={draw=none},
tick align=outside,
tick pos=left,
axis background/.style={fill=white},
axis x line*=bottom,
axis y line*=left,
width = \linewidth,
height = 0.55\linewidth,
x grid style={white!69.01961!black},
xtick style={color=black},
y grid style={white!69.01961!black},
ytick style={color=black},
enlargelimits=false,
xlabel={Number of base station antennas (\(M\))},
xmajorgrids,
ymajorgrids,
]

\addplot [color=colorULA]
  table[row sep=crcr]{%
1	-0.269301021792899\\
2	-0.664127876255103\\
3	-0.930806344339999\\
4	-1.2078464170776\\
5	-1.40377245875995\\
6	-1.59710872401002\\
7	-1.77723182099449\\
8	-1.93374630013144\\
9	-2.07650152088067\\
10	-2.22239294873762\\
11	-2.32643759581358\\
12	-2.42978863631953\\
13	-2.53479603684567\\
14	-2.64425219749546\\
15	-2.72720525676483\\
16	-2.76920228950072\\
17	-2.8474030638923\\
18	-2.92935366248123\\
19	-3.01405796656259\\
20	-3.09090941123328\\
21	-3.15628097254344\\
22	-3.21680752257166\\
23	-3.24126256678473\\
24	-3.27125917044479\\
25	-3.30275488572536\\
26	-3.33324734123171\\
27	-3.36501470224919\\
28	-3.38941988566111\\
29	-3.42384457322938\\
30	-3.46359136375726\\
31	-3.5029374526418\\
};
\addlegendentry{ULA}

\addplot [color=colorURA]
  table[row sep=crcr]{%
1	-0.56172692186394\\
2	-1.12358000514905\\
3	-1.48734840210199\\
4	-1.71657939158185\\
5	-1.90831948072822\\
6	-2.10834655632804\\
7	-2.32577252737606\\
8	-2.49863876981775\\
9	-2.46535149373773\\
10	-2.45424285545428\\
11	-2.4316229161177\\
12	-2.42408547172199\\
13	-2.43912967221125\\
14	-2.48558602616809\\
15	-2.55083577709924\\
16	-2.59886271258805\\
17	-2.60229112517169\\
18	-2.60844886052757\\
19	-2.59026231347417\\
20	-2.58698574641554\\
21	-2.59974017998845\\
22	-2.62395904829656\\
23	-2.64481979299461\\
24	-2.69244366300558\\
25	-2.71347400603201\\
26	-2.72735725029838\\
27	-2.72805791914366\\
28	-2.73805162643945\\
29	-2.75962693147518\\
30	-2.78293919899675\\
31	-2.82986254168331\\
};
\addlegendentry{URA}

\addplot [color=coloriid]
  table[row sep=crcr]{%
1	0\\
2	-1.50514997831991\\
3	-2.38560627359831\\
4	-3.01029995663981\\
5	-3.49485002168009\\
6	-3.89075625191822\\
7	-4.22549020007128\\
8	-4.51544993495972\\
9	-4.77121254719663\\
10	-5\\
11	-5.20696342579113\\
12	-5.39590623023812\\
13	-5.56971676153418\\
14	-5.73064017839119\\
15	-5.88045629527841\\
16	-6.02059991327962\\
17	-6.15224460689137\\
18	-6.27636252551653\\
19	-6.39376800476415\\
20	-6.50514997831991\\
21	-6.6110964736696\\
22	-6.71211340411103\\
23	-6.80863918008796\\
24	-6.90105620855803\\
25	-6.98970004336019\\
26	-7.07486673985409\\
27	-7.15681882079494\\
28	-7.2357901567111\\
29	-7.31198998949478\\
30	-7.38560627359831\\
31	-7.45680846917136\\
};
\addlegendentry{i.i.d.}

\end{axis}
\end{tikzpicture}%
    \caption{Average standard deviation of channel gain (\ref{eq:std}) when increasing the number of antennas for the ULA and URA.}%
    \label{fig:chhard}
\end{figure}

\textbf{Channel Hardening.} When increasing the number of antennas, the channel hardening effect appears. 
For a Rayleigh fading channel, the channel hardening becomes \(10 \log_{10}(\sqrt{M})\) when using the standard deviation for comparison as in \((\ref{eq:std})\). The channel hardening is here measured as the difference of the standard deviation of the channel gain when going from \num{1} to \(M\) antennas, i.e., \SI{7.5}{\deci\bel} with \num{31} antennas. A comparison between i.i.d. Rayleigh fading, the ULA and URA is shown in Fig.~\ref{fig:chhard}. Here, 
the average standard deviation of channel gain of a time window of size 600,
extracted from the continuous measurements along path B and C in Fig.~\ref{fig:map},
for an increasing number of antennas is depicted. The antennas are chosen in the order that is outlined in Fig.~\ref{fig:antenna-numbering}.

In Fig.~\ref{fig:chhard}, a clear channel hardening effect can be seen as the standard deviation of channel gain decreases when the number of base station antennas increases. 
In the beginning the URA has a lower standard deviation than the ULA until eight antennas where it saturates and is then bypassed by the ULA. For the ULA the average channel hardening is \SI{3.2}{\deci\bel} and for the URA it is \SI{2.3}{\deci\bel}. The reason for more channel hardening with the ULA is most likely due to better possibilities of exploiting the spatial diversity while adding more rows to the URA does not contribute as much. The channel hardening effect allows to reduce the fading margins and therefore the overall transmit power as well. 


\subsection{Serving Multiple Nodes.}\label{sec:serve-nodes}
The channel orthogonality is assessed by means of the correlation coefficient and the inverse condition number of channels captured at different measurement locations. The former describes the orthogonality of two nodes, while the latter shows the  
joint channel orthogonality of multiple nodes. Both metrics are evaluated with respect to the number of base station antennas.
Per antenna configuration, we have on average 300~locations. The actual static locations were extended with virtual locations by splitting the continuous measurements in virtual locations, each with 100~channel instances, equivalent to capturing \num{1}~second. 

\textbf{Correlation Coefficient.}\label{sec:corr-coeff}
The correlation coefficient relates to the concept of favorable propagation, which also quantifies the ability to separate channels. When there is favorable propagation~\cite{bjornsonBook}, the channel vectors are pair-wise orthogonal such that \(\hermconj{\vectr{h}_i}\vectr{h}_j\rightarrow 0\  \text{as}\   M\rightarrow \infty\) for two positions \(i\) and \(j\).


 \begin{figure}[h]
     \centering
\begin{tikzpicture}

\definecolor{colorULA}{rgb}{0.12157,0.46667,0.70588}
\definecolor{colorURA}{rgb}{1.00000,0.49804,0.05490}
\definecolor{coloriid}{rgb}{0.17255,0.62745,0.17255}

\begin{axis}[
xmin=-0.50000, xmax=32.50000,
ymin=-0.5, ymax=1.04839,
axis lines* = {left},
legend style={fill opacity=0.8, draw opacity=1, text opacity=1, draw=white!80.00000!black},
legend style={nodes={scale=0.9, transform shape}},
legend style={draw=none},
tick align=outside,
tick pos=left,
axis background/.style={fill=white},
axis x line*=bottom,
axis y line*=left,
width = \linewidth,
height = 0.5\linewidth,
x grid style={white!69.01961!black},
xtick style={color=black},
y grid style={white!69.01961!black},
ymode=log,
ytick style={color=black},
enlargelimits=false,
xlabel={Number of base station antennas (\(M\))},
ylabel={\(\delta_{i,j}\)},
xmajorgrids,
ymajorgrids,
]
\addplot [thick, colorULA]
table [row sep=\\] {%
1.00000 1.00000\\2.00000 0.51316\\3.00000 0.34766\\4.00000 0.26238\\5.00000 0.21301\\6.00000 0.17863\\7.00000 0.15495\\8.00000 0.13550\\9.00000 0.12105\\10.00000 0.10970\\11.00000 0.09975\\12.00000 0.09207\\13.00000 0.08476\\14.00000 0.07877\\15.00000 0.07425\\16.00000 0.06999\\17.00000 0.06577\\18.00000 0.06218\\19.00000 0.05899\\20.00000 0.05637\\21.00000 0.05379\\22.00000 0.05186\\23.00000 0.04955\\24.00000 0.04770\\25.00000 0.04560\\26.00000 0.04421\\27.00000 0.04241\\28.00000 0.04079\\29.00000 0.03957\\30.00000 0.03832\\31.00000 0.03709\\};
\addlegendentry{ULA}
\addplot [thick, colorURA]
table [row sep=\\] {%
1.00000 1.00000\\2.00000 0.52017\\3.00000 0.35939\\4.00000 0.27597\\5.00000 0.22443\\6.00000 0.18985\\7.00000 0.16531\\8.00000 0.14831\\9.00000 0.14772\\10.00000 0.14408\\11.00000 0.14042\\12.00000 0.13698\\13.00000 0.13377\\14.00000 0.12835\\15.00000 0.12507\\16.00000 0.12094\\17.00000 0.12034\\18.00000 0.11798\\19.00000 0.11467\\20.00000 0.11353\\21.00000 0.11078\\22.00000 0.10833\\23.00000 0.10515\\24.00000 0.10269\\25.00000 0.10161\\26.00000 0.09869\\27.00000 0.09699\\28.00000 0.09542\\29.00000 0.09256\\30.00000 0.09071\\31.00000 0.08889\\};
\addlegendentry{URA}
\addplot [thick, coloriid]
table [row sep=\\] {%
1.00000 1.00000\\2.00000 0.50000\\3.00000 0.33333\\4.00000 0.25000\\5.00000 0.20000\\6.00000 0.16667\\7.00000 0.14286\\8.00000 0.12500\\9.00000 0.11111\\10.00000 0.10000\\11.00000 0.09091\\12.00000 0.08333\\13.00000 0.07692\\14.00000 0.07143\\15.00000 0.06667\\16.00000 0.06250\\17.00000 0.05882\\18.00000 0.05556\\19.00000 0.05263\\20.00000 0.05000\\21.00000 0.04762\\22.00000 0.04545\\23.00000 0.04348\\24.00000 0.04167\\25.00000 0.04000\\26.00000 0.03846\\27.00000 0.03704\\28.00000 0.03571\\29.00000 0.03448\\30.00000 0.03333\\31.00000 0.03226\\};
\addlegendentry{i.i.d.}
\end{axis}

\end{tikzpicture}
     \caption{Average node correlation \(\delta_{i,j}\) in (\ref{eq:corr-coeff}) between the channel vectors of two random IoT nodes, as a function of the number of BS antennas. Expressed in \si{\deci\bel} to better illustrate the difference between the graphs.}%
     \label{fig:corr-coefficient}
\end{figure}
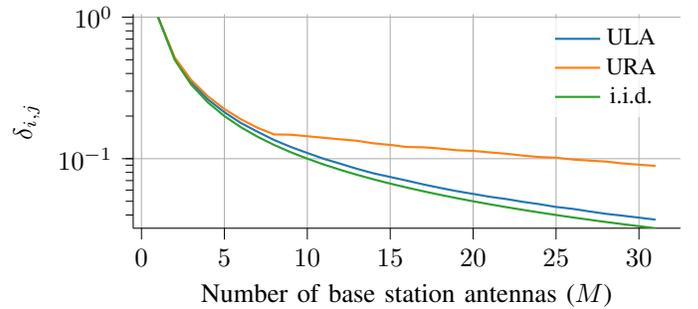


 Fig.~\ref{fig:corr-coefficient} depicts the correlation between two channel instances from two random locations as a function of the number of base station antennas. 
 The result was obtained by calculating the correlation coefficient -- as defined in (\ref{eq:corr-coeff}) -- \num{100 000} times per antenna configuration and for different numbers of consecutive antennas. For i.i.d. Rayleigh fading the average correlation coefficient becomes \(1/M\)~\cite{bjornsonBook}.
 Fig.~\ref{fig:corr-coefficient} shows the same trend for the ULA, URA and i.i.d. Rayleigh fading for the first eight antennas. After eight antennas the decrease of the correlation flattens for the URA, demonstrating that adding a second row does not contribute as much to the decorrelation of the channels. 
 This effect is not as noticeable for the ULA case, illustrating that increasing the size of the ULA increases the spatial diversity, as was also observed when studying the channel hardening. 

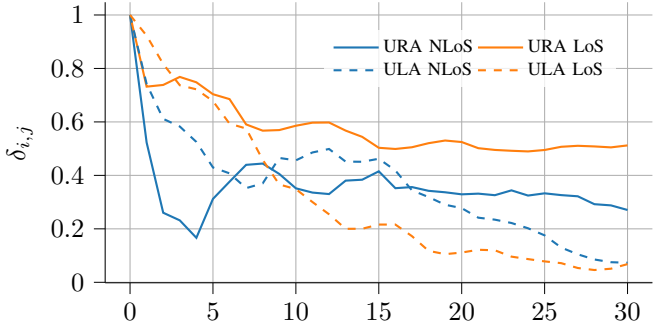
\begin{figure}[h]
     \centering
\begin{tikzpicture}

\definecolor{colorLoS}{rgb}{0.12157,0.46667,0.70588}
\definecolor{colorNLoS}{rgb}{1.00000,0.49804,0.05490}

\begin{axis}[
legend cell align={left},
 legend columns=2, 
axis lines* = {left},
legend style={fill opacity=0.8, draw opacity=1, text opacity=1, at={(0.4,0.9)}, anchor=north west, draw=white!80.00000!black},
legend style={nodes={scale=0.7, transform shape}},
legend style={draw=none},
tick align=outside,
tick pos=left,
height = 0.6\linewidth,
width = \linewidth,
x grid style={white!69.01961!black},
xmin=-1.50000, xmax=31.50000,
xtick style={color=black},
y grid style={white!69.01961!black},
ymin=-0.00173, ymax=1.04770,
ytick style={color=black},
ylabel={\(\delta_{i,j}\)},
tick align=outside,
axis background/.style={fill=white},
axis x line*=bottom,
axis y line*=left,
xmajorgrids,
ymajorgrids,
]
\addplot [thick, colorLoS]
table [row sep=\\] {%
0.00000 1.00000\\1.00000 0.52290\\2.00000 0.26022\\3.00000 0.23170\\4.00000 0.16648\\5.00000 0.31203\\6.00000 0.37637\\7.00000 0.43941\\8.00000 0.44423\\9.00000 0.40532\\10.00000 0.35161\\11.00000 0.33565\\12.00000 0.33011\\13.00000 0.38035\\14.00000 0.38380\\15.00000 0.41519\\16.00000 0.35207\\17.00000 0.35626\\18.00000 0.34224\\19.00000 0.33677\\20.00000 0.32909\\21.00000 0.33165\\22.00000 0.32588\\23.00000 0.34396\\24.00000 0.32453\\25.00000 0.33283\\26.00000 0.32618\\27.00000 0.32178\\28.00000 0.29225\\29.00000 0.28777\\30.00000 0.27056\\};
\addlegendentry{URA NLoS}

\addplot [thick, colorNLoS]
table [row sep=\\] {%
0.00000 1.00000\\1.00000 0.73192\\2.00000 0.73831\\3.00000 0.76832\\4.00000 0.74819\\5.00000 0.70359\\6.00000 0.68529\\7.00000 0.59066\\8.00000 0.56727\\9.00000 0.56947\\10.00000 0.58570\\11.00000 0.59716\\12.00000 0.59780\\13.00000 0.56752\\14.00000 0.54409\\15.00000 0.50319\\16.00000 0.49879\\17.00000 0.50523\\18.00000 0.52040\\19.00000 0.53037\\20.00000 0.52472\\21.00000 0.50173\\22.00000 0.49519\\23.00000 0.49202\\24.00000 0.48963\\25.00000 0.49500\\26.00000 0.50694\\27.00000 0.51065\\28.00000 0.50829\\29.00000 0.50475\\30.00000 0.51211\\};
\addlegendentry{URA LoS}
\addplot [thick, dashed, colorLoS]
table [row sep=\\] {%
0.00000 1.00000\\1.00000 0.74174\\2.00000 0.61135\\3.00000 0.58275\\4.00000 0.52504\\5.00000 0.42966\\6.00000 0.40762\\7.00000 0.35233\\8.00000 0.37015\\9.00000 0.46514\\10.00000 0.45620\\11.00000 0.48574\\12.00000 0.49877\\13.00000 0.45150\\14.00000 0.45088\\15.00000 0.46216\\16.00000 0.41831\\17.00000 0.34409\\18.00000 0.31889\\19.00000 0.29016\\20.00000 0.27741\\21.00000 0.24176\\22.00000 0.23434\\23.00000 0.22187\\24.00000 0.20216\\25.00000 0.17560\\26.00000 0.13002\\27.00000 0.10475\\28.00000 0.08538\\29.00000 0.07522\\30.00000 0.07321\\};
\addlegendentry{ULA NLoS}
\addplot [thick,dashed, colorNLoS]
table [row sep=\\] {%
0.00000 1.00000\\1.00000 0.92258\\2.00000 0.81662\\3.00000 0.73712\\4.00000 0.72191\\5.00000 0.67784\\6.00000 0.59337\\7.00000 0.57532\\8.00000 0.45797\\9.00000 0.36598\\10.00000 0.34865\\11.00000 0.30031\\12.00000 0.25448\\13.00000 0.20011\\14.00000 0.20034\\15.00000 0.21599\\16.00000 0.21607\\17.00000 0.17263\\18.00000 0.11773\\19.00000 0.10528\\20.00000 0.11134\\21.00000 0.12172\\22.00000 0.11976\\23.00000 0.09603\\24.00000 0.08683\\25.00000 0.07824\\26.00000 0.07164\\27.00000 0.05353\\28.00000 0.04597\\29.00000 0.05083\\30.00000 0.06806\\};
\addlegendentry{ULA LoS}
\end{axis}

\end{tikzpicture}
     \caption{Correlation coefficient \(\delta_{i,j}\) between the two LoS and NLoS users.}%
     \label{fig:corr-coefficient-los-vs-nlos}
\end{figure}

The correlation between the channels for a LoS and NLoS case is shown in Fig.~\ref{fig:corr-coefficient-los-vs-nlos}.
The general trend shows that the ULA configuration captures channels which are less correlated than the URA configuration. Moreover, when deploying close to \num{32} antennas, the correlation coefficient for both NLoS and LoS becomes equal. Hence, even in LoS, increasing the number of antenna elements in a ULA configuration contributes to the decorrelation of the channel coefficients such that nodes could be separated. However, for the URA, as expected this is trickier in LoS and the decorrelation of users is much more prominent in the NLoS scenario.

\textbf{Orthogonality of Multiple Nodes.}
The orthogonality of multiple IoT nodes is assessed by examining the condition number~\cite{6328480} or the distance from favorable propagation~\cite{6951994}.
 The normalization and the channel instance selection procedure are equivalent to that explained in Section~\ref{sec:corr-coeff}.

\begin{figure}[tbp]
    \centering
\begin{tikzpicture}

    \definecolor{color0}{rgb}{0.12157,0.46667,0.70588}
    \definecolor{color1}{rgb}{1.00000,0.49804,0.05490}
    \definecolor{color2}{rgb}{0.17255,0.62745,0.17255}
    
    \begin{axis}[
    axis lines* = {left},
    legend cell align={left},
    legend style={fill opacity=0.8, draw opacity=1, text opacity=1, at={(0.03000,0.97000)}, anchor=north west, draw=white!80.00000!black},
    legend style={nodes={scale=0.7, transform shape}},
    legend style={draw=none},
    tick align=outside,
    tick pos=left,
    width = \linewidth,
    height= 0.7\linewidth,
    x grid style={white!69.01961!black},
    xmin=-0.50000, xmax=40,
    xtick={0,5,10,15,20,25,30},
    xticklabels = {0,5,10,15,20,25,30},
    xtick style={color=black},
    y grid style={white!69.01961!black},
    ymin=-0.03658, ymax=1,
    ytick style={color=black},
    xlabel={Number of base station antennas (M)},
    ylabel={Average inverse condition number \(\kappa_{K,M}^{-1}\)},
    legend entries={ULA,
                URA,
                i.i.d.},
                tick align=outside,
axis background/.style={fill=white},
axis x line*=bottom,
axis y line*=left,
xmajorgrids,
ymajorgrids,
    ]
\addlegendimage{dashed ,color=black}
\addlegendimage{color=black}
\addlegendimage{dotted ,color=black}

\addplot [color0, dashed]
table [row sep=\\] {%
1.00000 0.00000\\2.00000 0.16611\\3.00000 0.26051\\4.00000 0.32544\\5.00000 0.37334\\6.00000 0.41293\\7.00000 0.44409\\8.00000 0.47154\\9.00000 0.49476\\10.00000 0.51603\\11.00000 0.53451\\12.00000 0.55164\\13.00000 0.56586\\14.00000 0.58003\\15.00000 0.59303\\16.00000 0.60469\\17.00000 0.61597\\18.00000 0.62649\\19.00000 0.63567\\20.00000 0.64504\\21.00000 0.65426\\22.00000 0.66346\\23.00000 0.67138\\24.00000 0.67991\\25.00000 0.68652\\26.00000 0.69478\\27.00000 0.70208\\28.00000 0.70958\\29.00000 0.71714\\30.00000 0.72365\\31.00000 0.73105\\};

\addplot [color1, dashed]
table [row sep=\\] {%
1.00000 0.00000\\2.00000 0.00000\\3.00000 0.00000\\4.00000 0.00000\\5.00000 0.01602\\6.00000 0.03507\\7.00000 0.05453\\8.00000 0.07365\\9.00000 0.09133\\10.00000 0.10892\\11.00000 0.12477\\12.00000 0.13987\\13.00000 0.15475\\14.00000 0.16809\\15.00000 0.18110\\16.00000 0.19297\\17.00000 0.20539\\18.00000 0.21609\\19.00000 0.22695\\20.00000 0.23704\\21.00000 0.24661\\22.00000 0.25586\\23.00000 0.26549\\24.00000 0.27396\\25.00000 0.28202\\26.00000 0.29058\\27.00000 0.29787\\28.00000 0.30511\\29.00000 0.31308\\30.00000 0.31968\\31.00000 0.32659\\};
\addplot [color2, dashed]
table [row sep=\\] {%
1.00000 0.00000\\2.00000 0.00000\\3.00000 0.00000\\4.00000 0.00000\\5.00000 0.00000\\6.00000 0.00000\\7.00000 0.00000\\8.00000 0.00000\\9.00000 0.00000\\10.00000 0.00307\\11.00000 0.00748\\12.00000 0.01260\\13.00000 0.01827\\14.00000 0.02411\\15.00000 0.03014\\16.00000 0.03616\\17.00000 0.04223\\18.00000 0.04826\\19.00000 0.05416\\20.00000 0.05993\\21.00000 0.06583\\22.00000 0.07124\\23.00000 0.07679\\24.00000 0.08206\\25.00000 0.08736\\26.00000 0.09243\\27.00000 0.09726\\28.00000 0.10209\\29.00000 0.10712\\30.00000 0.11166\\31.00000 0.11613\\};
\addplot [color0]
table [row sep=\\] {%
1.00000 0.00000\\2.00000 0.16329\\3.00000 0.25117\\4.00000 0.31084\\5.00000 0.35388\\6.00000 0.38790\\7.00000 0.41411\\8.00000 0.43736\\9.00000 0.45708\\10.00000 0.47534\\11.00000 0.48957\\12.00000 0.50162\\13.00000 0.51412\\14.00000 0.52451\\15.00000 0.53555\\16.00000 0.54468\\17.00000 0.55175\\18.00000 0.56049\\19.00000 0.56823\\20.00000 0.57546\\21.00000 0.58127\\22.00000 0.58769\\23.00000 0.59416\\24.00000 0.60075\\25.00000 0.60543\\26.00000 0.61014\\27.00000 0.61448\\28.00000 0.61951\\29.00000 0.62481\\30.00000 0.63003\\31.00000 0.63470\\};

\addplot [color1]
table [row sep=\\] {%
1.00000 0.00000\\2.00000 0.00000\\3.00000 0.00000\\4.00000 0.00000\\5.00000 0.01099\\6.00000 0.02379\\7.00000 0.03599\\8.00000 0.04761\\9.00000 0.05841\\10.00000 0.06793\\11.00000 0.07698\\12.00000 0.08472\\13.00000 0.09203\\14.00000 0.09891\\15.00000 0.10501\\16.00000 0.11121\\17.00000 0.11627\\18.00000 0.12150\\19.00000 0.12590\\20.00000 0.13046\\21.00000 0.13424\\22.00000 0.13787\\23.00000 0.14150\\24.00000 0.14473\\25.00000 0.14813\\26.00000 0.15077\\27.00000 0.15379\\28.00000 0.15633\\29.00000 0.15947\\30.00000 0.16136\\31.00000 0.16421\\};
\addplot [color2]
table [row sep=\\] {%
1.00000 0.00000\\2.00000 0.00000\\3.00000 0.00000\\4.00000 0.00000\\5.00000 0.00000\\6.00000 0.00000\\7.00000 0.00000\\8.00000 0.00000\\9.00000 0.00000\\10.00000 0.00113\\11.00000 0.00267\\12.00000 0.00443\\13.00000 0.00626\\14.00000 0.00810\\15.00000 0.00993\\16.00000 0.01167\\17.00000 0.01337\\18.00000 0.01500\\19.00000 0.01648\\20.00000 0.01794\\21.00000 0.01932\\22.00000 0.02062\\23.00000 0.02197\\24.00000 0.02320\\25.00000 0.02429\\26.00000 0.02522\\27.00000 0.02630\\28.00000 0.02735\\29.00000 0.02833\\30.00000 0.02917\\31.00000 0.02997\\};
\addplot [color0, dotted]
table [row sep=\\] {%
1.00000 0.00000\\2.00000 0.22705\\3.00000 0.33287\\4.00000 0.39919\\5.00000 0.44754\\6.00000 0.48424\\7.00000 0.51297\\8.00000 0.53729\\9.00000 0.55805\\10.00000 0.57594\\11.00000 0.59167\\12.00000 0.60628\\13.00000 0.61778\\14.00000 0.62874\\15.00000 0.63878\\16.00000 0.64830\\17.00000 0.65775\\18.00000 0.66458\\19.00000 0.67177\\20.00000 0.67928\\21.00000 0.68509\\22.00000 0.69143\\23.00000 0.69709\\24.00000 0.70212\\25.00000 0.70708\\26.00000 0.71264\\27.00000 0.71564\\28.00000 0.72106\\29.00000 0.72575\\30.00000 0.72882\\31.00000 0.73263\\};

\addplot [color1, dotted]
table [row sep=\\] {%
1.00000 0.00000\\2.00000 0.00000\\3.00000 0.00000\\4.00000 0.00000\\5.00000 0.02021\\6.00000 0.04320\\7.00000 0.06628\\8.00000 0.08793\\9.00000 0.10825\\10.00000 0.12737\\11.00000 0.14541\\12.00000 0.16160\\13.00000 0.17724\\14.00000 0.19192\\15.00000 0.20525\\16.00000 0.21862\\17.00000 0.23079\\18.00000 0.24222\\19.00000 0.25329\\20.00000 0.26323\\21.00000 0.27346\\22.00000 0.28286\\23.00000 0.29196\\24.00000 0.30091\\25.00000 0.30892\\26.00000 0.31683\\27.00000 0.32481\\28.00000 0.33201\\29.00000 0.33896\\30.00000 0.34626\\31.00000 0.35237\\};
\addplot [color2, dotted]
table [row sep=\\] {%
1.00000 0.00000\\2.00000 0.00000\\3.00000 0.00000\\4.00000 0.00000\\5.00000 0.00000\\6.00000 0.00000\\7.00000 0.00000\\8.00000 0.00000\\9.00000 0.00000\\10.00000 0.00380\\11.00000 0.00922\\12.00000 0.01556\\13.00000 0.02242\\14.00000 0.02967\\15.00000 0.03702\\16.00000 0.04436\\17.00000 0.05188\\18.00000 0.05919\\19.00000 0.06644\\20.00000 0.07356\\21.00000 0.08073\\22.00000 0.08760\\23.00000 0.09431\\24.00000 0.10108\\25.00000 0.10745\\26.00000 0.11392\\27.00000 0.12013\\28.00000 0.12623\\29.00000 0.13198\\30.00000 0.13782\\31.00000 0.14365\\};
\node[color0] at (35,0.7) {\(K=2\)};
\node[color1] at (35,0.25) {\(K=5\)};
\node[color2] at (35,0.08) {\(K=10\)};
\end{axis}


\end{tikzpicture}
    \caption{The average inverse condition number \(\kappa_{K,M}^{-1}\) for \(K\) IoT nodes and \num{32} antennas.}
    \label{fig:condition-number}
\end{figure}
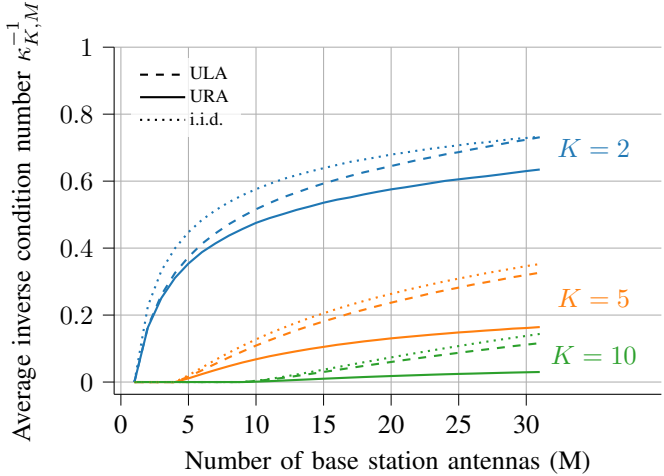

A large condition number implies strongly correlated channels, while a condition number of one indicates pair-wise orthogonal channels. The inverse condition number is chosen as a metric to be able to express the orthogonality in a finite range, i.e., \(\kappa_{K,M}^{-1} \in [0,1]\). 
The inverse condition number for two, five and ten users as a function of the number of base station is shown in Fig.~\ref{fig:condition-number}. The number of users are randomly selected as discussed in Section~\ref{sec:corr-coeff}. Logically \(\kappa_{K,M}^{-1}\) equals zero as long as \(K\) is larger than \(M\). 

We see that the \(\kappa_{K,M}^{-1}\) of the ULA follows closely the ideal i.i.d. Rayleigh fading case and even tends to move closer to this curve as the number of base station antennas increases; for the URA case, the distance from i.i.d Rayleigh fading is larger.  


\begin{figure}[tbp]
    \centering
    \input{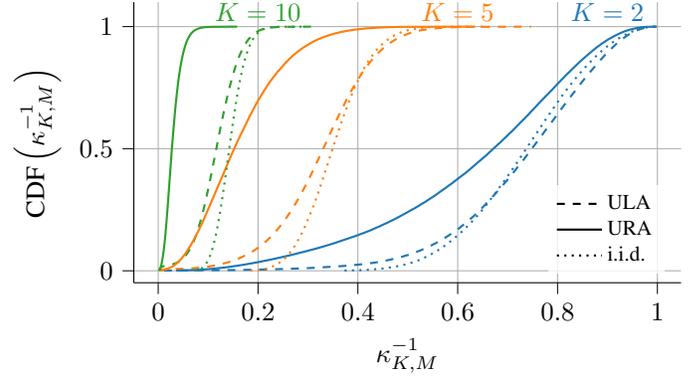}
    \caption{CDF of the inverse condition number \(\kappa_{K,M}^{-1}\) for \(K\) IoT nodes and \num{32} antennas.}
    \label{fig:condition-number-cdf}
\end{figure}

The empirical CDF is used to further elaborate on the impact of \(K\) and \(M\) in Fig~\ref{fig:condition-number-cdf}. There exist some strongly correlated channels as is illustrated by the spread of \(\kappa_{K,M}^{-1}\) over the entire range \([0,1]\). This is also noticeable in the CDF of the correlation coefficient (Fig.~\ref{fig:condition-number-cdf}) as both ULA and URA show some correlation coefficients close to one.
While there are not many strongly correlated signals, it still demonstrates the importance of adequate user grouping and scheduling.

\subsection{Scheduling IoT Nodes}\label{sec:schedule-nodes}

\textbf{Dominant Eigendirections.}
As demonstrated with previous metrics, we cannot assume i.i.d. Rayleigh fading channels but expect that signals, coming from the IoT nodes, have distinct directions. 
In other words, we expect that in this setting, nodes have dominant directions which do not change drastically over time. These dominant directions and their significance are studied by their eigendirections and could be used to schedule nodes.

The distribution of weak and strong eigendirections can be extracted from the channel correlation matrix $\matr{R}$ as defined in (\ref{eq.R})~\cite{bjornsonBook}. Similar as the channel hardening analysis, the correlation matrix is obtained from the continuous channel collection along path B and C for a time window of size \num{600} samples, corresponding to approximately \(25 \lambda\).
Through the eigenvalue decomposition of the correlation matrix, a diagonal matrix $\matr{D}$ containing the eigenvalues is obtained.  The eigenvalues are sorted in descending order. The higher the values of the first eigenvalues, the more energy is confined in a few eigendirections. Fig.~\ref{fig:eigenvaluesAB} depicts the eigenvalues  $\operatorname{tr}(\matr{D})$, sorted in descending order, of the channels for path C and B in Fig.~\ref{fig:los-nlos-points}. Most of the energy -- between 68\% and 85\% -- is carried by three eigendirections, regardless of the antenna array and node position. 
As we have a finite number of measured channel instances, the i.i.d. Rayleigh fading channel is also simulated with the same number of snapshots. As can be observed in Fig.~\ref{fig:eigenvaluesAB}, the eigenvalues are distributed around \SI{0}{\deci\bel}, while asymptotically all eigenvalues will be equal; Fig.~\ref{fig:eigenvaluesAB} shows that we are close to the asymptotic case. We can conclude that for both configurations and both paths, dominant directions are present; this in contrast to the simulated i.i.d. case. Also, the ULA and URA show similar trends on both paths.



\begin{figure}[tbp]
    \centering
%
%
\definecolor{color1}{rgb}{0.12157,0.46667,0.70588}
\definecolor{color2}{rgb}{1.00000,0.49804,0.05490}
\begin{tikzpicture}

\begin{axis}[%
width=0.8\linewidth,
height=0.3\linewidth,
scale only axis,
xmin=-0.1,
xmax=32,
xlabel={Number of base station antennas \(M\)},
ymin=-0.5,
ymax=6.1,
ylabel={Chordal distance},
axis background/.style={fill=white},
tick align=outside,
axis background/.style={fill=white},
axis x line*=bottom,
axis y line*=left,
xmajorgrids,
ymajorgrids,
axis lines* = {left},
    legend cell align={left},
    legend style={fill opacity=0.8, draw opacity=1, text opacity=1, at={(0.5,0.4)}, anchor=north west, draw=white!80.00000!black},
    legend style={nodes={scale=0.7, transform shape}},
    legend style={draw=none},
    tick align=outside,
    tick pos=left
]
\addplot [color=color1]
  table[row sep=crcr]{%
1	0\\
2	5.38155094492253e-31\\
3	1.61254111391657e-30\\
4	1.37019302525316\\
5	2.1742515067223\\
6	2.78722668292652\\
7	3.13507258100883\\
8	3.51593078345989\\
9	4.08034647846343\\
10	4.30334040687303\\
11	4.38103889436627\\
12	4.67645897946375\\
13	4.81584432224735\\
14	4.84494612000858\\
15	4.92035894476257\\
16	5.02353331523317\\
17	5.0476669726643\\
18	5.06841945218631\\
19	5.1466542161809\\
20	5.17113913290531\\
21	5.15872549913484\\
22	5.21083286135548\\
23	5.2529714736289\\
24	5.28088920967136\\
25	5.3062839637944\\
26	5.34321024538553\\
27	5.35207702340451\\
28	5.3742533903509\\
29	5.41721563759303\\
30	5.41124400114469\\
31	5.46746307284011\\
};
\addlegendentry{ULA}


\addplot [color=color2]
  table[row sep=crcr]{%
1	0\\
2	7.8926074189735e-31\\
3	1.25071931518944e-30\\
4	1.3424967650409\\
5	2.21776422119884\\
6	2.89806369445393\\
7	3.43893182446219\\
8	3.81236501371685\\
9	3.86806470099771\\
10	3.97807305087271\\
11	4.08627709563337\\
12	4.09109418684431\\
13	4.10348139596179\\
14	4.13113162565609\\
15	4.20223940488884\\
16	4.1738168839605\\
17	4.23431560767464\\
18	4.31151607426885\\
19	4.39960501455563\\
20	4.39499285981111\\
21	4.42309548332064\\
22	4.43602963168321\\
23	4.45118909656643\\
24	4.46159710558622\\
25	4.5350584974676\\
26	4.60034312117101\\
27	4.66768044081428\\
28	4.69870555030804\\
29	4.73295007907129\\
30	4.77030745821224\\
31	4.8254219380161\\
};
\addlegendentry{URA}


\end{axis}

\end{tikzpicture}%
    \caption{Chordal distance of 3-dominant eigenspaces between path B and C for the ULA and URA configuration.}
    \label{fig:chordal}
\end{figure}
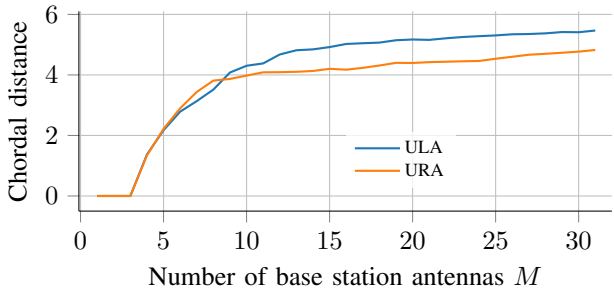

The difference between the three dominant eigendirections is measured by computing the chordal distance, defined in (\ref{eq.chordal}). The chordal distance is used in~\cite{5763372, 6831692} to reduce the complexity when selecting users to be scheduled together, while still performing similar to full-CSI selection algorithms. Fig.~\ref{fig:chordal} presents the average chordal distance between the 3-dominant eigenspaces of paths B and C. This figure shows the degree of orthogonality of the eigenspaces are between the aforementioned configurations. 
As expected, the difference between the eigenspaces increase when the number of antennas increases.
As also observed when studying the previous metrics, the ULA increases the distance between the three eigenspace more rapidly by increasing the number of base station antennas, while with the URA it stagnates after \num{8} antennas.

\begin{figure}[tbp]
    \centering
    \begin{tikzpicture}

\begin{axis}[%
scale only axis,
legend cell align={left},
tick align=outside,
tick pos=left,
x grid style={white!69.01961!black},
xmin=-0.1,
xmax=33.1,
xlabel={Eigenvalue number in decreasing order},
ymin=-25.1,
ymax=20.1,
axis lines* = {left},
ylabel={Eigenvalue [\si{\deci\bel}]},
legend style={fill opacity=0.8, draw opacity=1, text opacity=1, at={(0.15,0.95)}, anchor=north west, draw=white!80.00000!black},
legend style={nodes={scale=0.7, transform shape}},
legend style={draw=none},
width=0.8\linewidth,
height = 0.5\linewidth,
tick align=outside,
axis background/.style={fill=white},
axis x line*=bottom,
axis y line*=left,
legend columns = 3,
xmajorgrids,
ymajorgrids,
]


\addplot [color=colorULA, dashed, line width=1.0pt]
  table[row sep=crcr]{%
1	11.728188666232\\
2	8.06239784592052\\
3	5.3326151773526\\
4	2.89250682338931\\
5	0.703440954943015\\
6	-0.831319940133273\\
7	-2.5633302762741\\
8	-3.95507960630791\\
9	-5.29769111578405\\
10	-6.64409569378057\\
11	-7.81284009724524\\
12	-8.60738742121004\\
13	-9.92765880370019\\
14	-10.8828185782297\\
15	-12.0401403170641\\
16	-13.0937775301192\\
17	-13.9529724898163\\
18	-14.674139470775\\
19	-15.3648879666252\\
20	-16.0151104315816\\
21	-16.5039291351644\\
22	-17.0691715697645\\
23	-17.501445695893\\
24	-18.0082746562339\\
25	-18.4167047436132\\
26	-18.7742376207734\\
27	-19.2026555318637\\
28	-19.5487976847343\\
29	-19.9108001913767\\
30	-20.2666115202132\\
31	-20.5531008981912\\
};
\addlegendentry{ULA B-path}

\addplot [color=colorURA, dashed, line width=1.0pt]
  table[row sep=crcr]{%
1	11.2976812430779\\
2	6.80251522352883\\
3	4.58513557589354\\
4	2.92355338898795\\
5	1.57435844878541\\
6	0.581171503401604\\
7	-0.342952954417626\\
8	-1.21330485511761\\
9	-2.09600091213307\\
10	-2.8747884338701\\
11	-3.77492207071343\\
12	-4.61384583587966\\
13	-5.48701325026824\\
14	-6.13728493681973\\
15	-6.91077098512224\\
16	-7.79884439288565\\
17	-8.5976489993027\\
18	-9.28772075632433\\
19	-10.0241854272464\\
20	-10.7679743350099\\
21	-11.6370855439219\\
22	-12.2208987819746\\
23	-13.0023707038378\\
24	-13.6606437451745\\
25	-14.3880893175466\\
26	-14.9442145096254\\
27	-15.4937731595051\\
28	-16.0874652008201\\
29	-16.6478947591534\\
30	-17.2022792521231\\
31	-17.7107932557599\\
};
\addlegendentry{ULA C-path}

\addplot[color=colorIID]
  table[row sep=crcr]{%
1	1.68276320768211\\
2	1.4545907675157\\
3	1.33891577259514\\
4	1.23844976167999\\
5	1.10221844035846\\
6	1.04596060831465\\
7	0.841704447279851\\
8	0.802794801237915\\
9	0.67081938163708\\
10	0.486170514029512\\
11	0.356691520252378\\
12	0.262907233662837\\
13	0.16635959197346\\
14	0.113656547009372\\
15	0.0222974529157562\\
16	-0.0862065861273119\\
17	-0.128378076976361\\
18	-0.258732381369641\\
19	-0.339333898634984\\
20	-0.459361592641013\\
21	-0.604851871832778\\
22	-0.771477815248845\\
23	-0.868622242169879\\
24	-0.976627572620822\\
25	-1.0435932328166\\
26	-1.21917258768534\\
27	-1.33338036947623\\
28	-1.54613564493426\\
29	-1.73097250355116\\
30	-1.92092328600178\\
31	-2.06555137253398\\
};

\addlegendentry{i.i.d.}

\addplot [color=colorULA, line width=1.0pt]
  table[row sep=crcr]{%
1	12.9899804385129\\
2	6.81429254506856\\
3	2.20950937514985\\
4	-0.315808783765816\\
5	-2.10893241423588\\
6	-3.58432120422611\\
7	-4.78623353989803\\
8	-5.94579030050971\\
9	-6.9003223219859\\
10	-7.5613170029341\\
11	-8.19713021774037\\
12	-8.70692750466312\\
13	-9.19625653979486\\
14	-9.57842352644622\\
15	-9.9707332533732\\
16	-10.245593025816\\
17	-10.5337619556618\\
18	-10.7800953157826\\
19	-11.0159297074843\\
20	-11.2364182679719\\
21	-11.4383794094803\\
22	-11.6598862801543\\
23	-11.8011699236263\\
24	-12.0080204920867\\
25	-12.1778931061097\\
26	-12.3646496550312\\
27	-12.5568116908697\\
28	-12.754179873189\\
29	-12.9523524865481\\
30	-13.1751671990004\\
31	-13.4153236524575\\
};
\addlegendentry{URA B-path}

\addplot [color=colorURA, line width=1.0pt]
  table[row sep=crcr]{%
1	12.4532390366432\\
2	6.48501474972494\\
3	3.63550462935909\\
4	2.14112169614612\\
5	0.316151925243641\\
6	-0.942868623017083\\
7	-2.22514651345742\\
8	-3.58760686709554\\
9	-4.56101523292847\\
10	-5.6622566580704\\
11	-6.67422225357785\\
12	-7.70258892036029\\
13	-8.64192028693605\\
14	-9.40725150026656\\
15	-10.111049348693\\
16	-10.7652090968699\\
17	-11.3581782064755\\
18	-11.8032655748548\\
19	-12.2654072782847\\
20	-12.6999041809362\\
21	-13.0651533650872\\
22	-13.3882185924001\\
23	-13.690307443643\\
24	-13.9773966449823\\
25	-14.2509992374388\\
26	-14.4868681729815\\
27	-14.7358144123415\\
28	-14.9852331660537\\
29	-15.2427718543441\\
30	-15.5159793519325\\
31	-15.753964751571\\
};
\addlegendentry{URA C-path}

\end{axis}

\end{tikzpicture}%
    \caption{Eigen-values ordered in descending order for different antenna topologies and different positions.}
    \label{fig:eigenvaluesAB}
\end{figure}
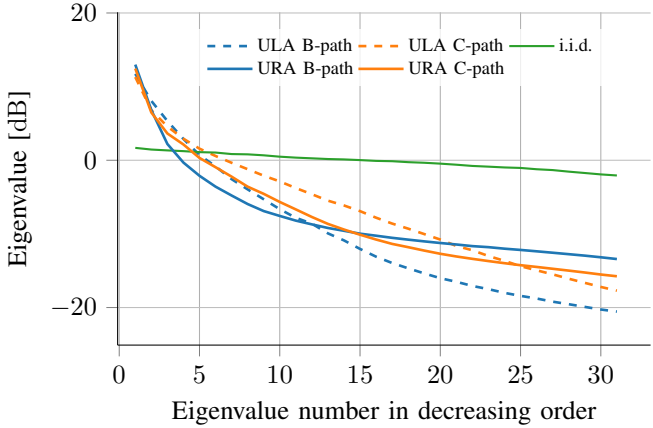

\section{Outlook and Conclusion}\label{sec:concl}
In this paper we have studied the opportunities of massive MIMO technology to upgrade LPWANs operating in the unlicensed 
sub-GHz
band for future IoT applications. We have evaluated several relevant performance metrics, based on measurement data gathered through an experimental campaign; this data is now available open-source. The campaign covered both LoS and NLoS scenarios, and ULA and URA antenna array configurations were deployed.
Our assessment confirms that both array gain and channel hardening are considerable. Hence, when combining the many antennas properly, the power consumption of the nodes can be reduced drastically and the link reliability can be increased. For example, for the case of a 32-antenna gateway this could specifically allow the battery-powered IoT devices to reduce their transmit power by \(M\) -- or \(10 \log_{10}(M)\) \si{\deci\bel}. Or alternatively, use the additional combining gain to extend the coverage. 
Furthermore, the fading margins can be lowered due to the channel hardening effect. 
This effect also ensures that less retransmissions are required, which are shown in~\cite{callebaut2019cross} to account for considerable energy consumption in current LPWAN technologies.



We performed an analysis of joint orthogonality of channels for different node positions. It shows that the relatively large array does receive quite different responses from these positions, opening the opportunity for spatial multiplexing of nodes. When comparing the URA and ULA, it is noticeable that while the received gains of both are similar, the channel correlation between different positions are quite different. This indicates that the average channel power per antenna is similar for both configurations, but the linear array also ensures a more diverse set of paths is received. The latter facilitates better decorrelation of nodes and is hence better suited for serving many nodes.
Noteworthy is that even with users perpendicular to the array, the ULA still performs better than the URA. The ULA also outperforms the URA when users have the same azimuth angle with respect to the base station, as shown in Fig.~\ref{fig:corr-coefficient-los-vs-nlos}.

The measured channels show the presence of a few dominant eigendirections that could used for scheduling users. It also demonstrates that i.i.d. Rayleigh fading does not hold in our scenarios. New models need to be designed to include sub-GHz massive MIMO propagation. These models can then be used to design new low-power IoT protocols. 







\section{Acknowledgment}
This work was funded by the European Union's Horizon 2020 under grant agreement no.~732174 (ORCA project) and no.~731884 (IoF2020 program - IoTrailer use case). We would like to thank Vladimir Volski for designing the path antenna, Sofie Pollin for her help with the Massive MIMO testbed configuration, François Rottenberg for the fruitful discussions and our colleagues of Dramco for assisting us during the experiments.

 \FloatBarrier
\footnotesize
\bibliographystyle{IEEEtranN}
{
\bibliography{bib}}

\begin{thebibliography}{32}
\providecommand{\natexlab}[1]{#1}
\providecommand{\url}[1]{#1}
\csname url@samestyle\endcsname
\providecommand{\newblock}{\relax}
\providecommand{\bibinfo}[2]{#2}
\providecommand{\BIBentrySTDinterwordspacing}{\spaceskip=0pt\relax}
\providecommand{\BIBentryALTinterwordstretchfactor}{4}
\providecommand{\BIBentryALTinterwordspacing}{\spaceskip=\fontdimen2\font plus
\BIBentryALTinterwordstretchfactor\fontdimen3\font minus
  \fontdimen4\font\relax}
\providecommand{\BIBforeignlanguage}[2]{{%
\expandafter\ifx\csname l@#1\endcsname\relax
\typeout{** WARNING: IEEEtranN.bst: No hyphenation pattern has been}%
\typeout{** loaded for the language `#1'. Using the pattern for}%
\typeout{** the default language instead.}%
\else
\language=\csname l@#1\endcsname
\fi
#2}}
\providecommand{\BIBdecl}{\relax}
\BIBdecl

\bibitem[{Callebaut} and {Van der Perre}(2020)]{cal2020}
G.~{Callebaut} and L.~{Van der Perre}, ``{Characterization of LoRa
  Point-to-Point Path Loss: Measurement Campaigns and Modeling Considering
  Censored Data},'' \emph{IEEE Internet of Things Journal}, vol.~7, no.~3, pp.
  1910--1918, 2020.

\bibitem[Xhonneux et~al.(2020)Xhonneux, Tapparel, Afisiadis,
  Balatsoukas-Stimming, and Burg]{matthieuAsilomar}
M.~Xhonneux, J.~Tapparel, O.~Afisiadis, A.~Balatsoukas-Stimming, and A.~Burg,
  ``{A Maximum-Likelihood-based Multi-User LoRa Receiver Implemented in GNU
  Radio},'' in \emph{2020 54rd Asilomar Conference on Signals, Systems, and
  Computers}, Nov. 2020.

\bibitem[Mahfoudi et~al.(2019)Mahfoudi, Sivadoss, Korachi, Turletti, and
  Dabbous]{doi:10.1002/itl2.120}
\BIBentryALTinterwordspacing
M.~N. Mahfoudi, G.~Sivadoss, O.~B. Korachi, T.~Turletti, and W.~Dabbous,
  ``{Joint Range Extension and Localization for Low-Power Wide-Area Network},''
  \emph{Internet Technology Letters}, vol.~2, no.~5, p. e120, 2019. [Online].
  Available: \url{https://onlinelibrary.wiley.com/doi/abs/10.1002/itl2.120}
\BIBentrySTDinterwordspacing

\bibitem[{Dongare} et~al.(2018){Dongare}, {Narayanan}, {Gadre}, {Luong},
  {Balanuta}, {Kumar}, {Iannucci}, and {Rowe}]{8480036}
A.~{Dongare}, R.~{Narayanan}, A.~{Gadre}, A.~{Luong}, A.~{Balanuta},
  S.~{Kumar}, B.~{Iannucci}, and A.~{Rowe}, ``{Charm: Exploiting Geographical
  Diversity through Coherent Combining in Low-Power Wide-Area Networks},'' in
  \emph{2018 17th ACM/IEEE International Conference on Information Processing
  in Sensor Networks (IPSN)}, 2018, pp. 60--71.

\bibitem[{Hoeller} et~al.(2018){Hoeller}, {Souza}, {Alcaraz López}, {Alves},
  {de Noronha Neto}, and {Brante}]{8372906}
A.~{Hoeller}, R.~D. {Souza}, O.~L. {Alcaraz López}, H.~{Alves}, M.~{de Noronha
  Neto}, and G.~{Brante}, ``{Analysis and Performance Optimization of LoRa
  Networks With Time and Antenna Diversity},'' \emph{IEEE Access}, vol.~6, pp.
  32\,820--32\,829, 2018.

\bibitem[Bj\"{o}rnson et~al.(2017{\natexlab{a}})Bj\"{o}rnson, {de Carvalho},
  {Sorensen}, {Larsson}, and {Popovski}]{7878690}
E.~Bj\"{o}rnson, E.~{de Carvalho}, J.~H. {Sorensen}, E.~G. {Larsson}, and
  P.~{Popovski}, ``{A Random Access Protocol for Pilot Allocation in Crowded
  Massive MIMO Systems},'' \emph{IEEE Transactions on Wireless Communications},
  vol.~16, no.~4, pp. 2220--2234, 2017.

\bibitem[Bana et~al.(2019)Bana, {de Carvalho}, Soret, Abrão, Marinello,
  Larsson, and Popovski]{BANA2019100859}
\BIBentryALTinterwordspacing
A.-S. Bana, E.~{de Carvalho}, B.~Soret, T.~Abrão, J.~C. Marinello, E.~G.
  Larsson, and P.~Popovski, ``{Massive MIMO for Internet of Things (IoT)
  connectivity},'' \emph{Physical Communication}, vol.~37, p. 100859, 2019.
  [Online]. Available:
  \url{http://www.sciencedirect.com/science/article/pii/S1874490719303891}
\BIBentrySTDinterwordspacing

\bibitem[{Lee} and {Yang}(2018)]{8241857}
B.~M. {Lee} and H.~{Yang}, ``{Massive MIMO for Industrial Internet of Things in
  Cyber-Physical Systems},'' \emph{IEEE Transactions on Industrial
  Informatics}, vol.~14, no.~6, pp. 2641--2652, 2018.

\bibitem[{Liu} and {Yu}(2018)]{8323218}
L.~{Liu} and W.~{Yu}, ``{Massive Connectivity With Massive MIMO—Part I:
  Device Activity Detection and Channel Estimation},'' \emph{IEEE Transactions
  on Signal Processing}, vol.~66, no.~11, pp. 2933--2946, 2018.

\bibitem[{Saxena} et~al.(2017){Saxena}, {Roy}, {Sahu}, and {Kim}]{7842418}
N.~{Saxena}, A.~{Roy}, B.~J.~R. {Sahu}, and H.~{Kim}, ``{Efficient IoT Gateway
  over 5G Wireless: A New Design with Prototype and Implementation Results},''
  \emph{IEEE Communications Magazine}, vol.~55, no.~2, pp. 97--105, 2017.

\bibitem[{Lee}(2018)]{8400514}
B.~M. {Lee}, ``{Improved Energy Efficiency of Massive MIMO-OFDM in
  Battery-Limited IoT Networks},'' \emph{IEEE Access}, vol.~6, pp.
  38\,147--38\,160, 2018.

\bibitem[Senel and Larsson(2018)]{senel2018grant}
K.~Senel and E.~G. Larsson, ``{Grant-free massive MTC-enabled massive MIMO: A
  compressive sensing approach},'' \emph{IEEE Transactions on Communications},
  vol.~66, no.~12, pp. 6164--6175, 2018.

\bibitem[{Ngo} et~al.(2014){Ngo}, {Larsson}, and {Marzetta}]{6951994}
H.~Q. {Ngo}, E.~G. {Larsson}, and T.~L. {Marzetta}, ``{Aspects of favorable
  propagation in Massive MIMO},'' in \emph{2014 22nd European Signal Processing
  Conference (EUSIPCO)}, 2014, pp. 76--80.

\bibitem[Gunnarsson et~al.(2020)Gunnarsson, Van~der Perre, and
  Tufvesson]{doi:10.1002/9781119471509.w5GRef040}
\BIBentryALTinterwordspacing
S.~Gunnarsson, L.~Van~der Perre, and F.~Tufvesson, \emph{Massive MIMO
  Channels}.\hskip 1em plus 0.5em minus 0.4em\relax American Cancer Society,
  2020, pp. 1--21. [Online]. Available:
  \url{https://onlinelibrary.wiley.com/doi/abs/10.1002/9781119471509.w5GRef040}
\BIBentrySTDinterwordspacing

\bibitem[{Gunnarsson} et~al.(2020){Gunnarsson}, {Flordelis}, {Van der Perre},
  and {Tufvesson}]{9069181}
S.~{Gunnarsson}, J.~{Flordelis}, L.~{Van der Perre}, and F.~{Tufvesson},
  ``{Channel Hardening in Massive MIMO: Model Parameters and Experimental
  Assessment},'' \emph{IEEE Open Journal of the Communications Society},
  vol.~1, pp. 501--512, 2020.

\bibitem[{Gao} et~al.(2015{\natexlab{a}}){Gao}, {Edfors}, {Rusek}, and
  {Tufvesson}]{7062910}
X.~{Gao}, O.~{Edfors}, F.~{Rusek}, and F.~{Tufvesson}, ``{Massive MIMO
  Performance Evaluation Based on Measured Propagation Data},'' \emph{IEEE
  Transactions on Wireless Communications}, vol.~14, no.~7, pp. 3899--3911,
  2015.

\bibitem[{Gao} et~al.(2015{\natexlab{b}}){Gao}, {Edfors}, {Tufvesson}, and
  {Larsson}]{7172496}
X.~{Gao}, O.~{Edfors}, F.~{Tufvesson}, and E.~G. {Larsson}, ``{Massive MIMO in
  Real Propagation Environments: Do All Antennas Contribute Equally?}''
  \emph{IEEE Transactions on Communications}, vol.~63, no.~11, pp. 3917--3928,
  2015.

\bibitem[{Arnold} and {ten Brink}(2019)]{8727203}
M.~{Arnold} and S.~{ten Brink}, ``{Properties of Measured Massive MIMO Channels
  using Different Antenna Geometries},'' in \emph{WSA 2019; 23rd International
  ITG Workshop on Smart Antennas}, 2019, pp. 1--5.

\bibitem[{Tran} et~al.(2019){Tran}, {Gustafsson}, {K\"{a}llström}, {Senel},
  and {Larsson}]{9048663}
M.~{Tran}, O.~{Gustafsson}, P.~{K\"{a}llström}, K.~{Senel}, and E.~G.
  {Larsson}, ``{An Architecture for Grant-Free Massive MIMO MTC Based on
  Compressive Sensing},'' in \emph{2019 53rd Asilomar Conference on Signals,
  Systems, and Computers}, 2019, pp. 901--905.

\bibitem[Beyene et~al.(2015)Beyene, Boyd, Ruttik, Bockelmann, Tirkkonen, and
  J{\"a}ntti]{beyene2015compressive}
Y.~Beyene, C.~Boyd, K.~Ruttik, C.~Bockelmann, O.~Tirkkonen, and R.~J{\"a}ntti,
  ``{Compressive sensing for MTC in new LTE uplink multi-user random access
  channel},'' in \emph{AFRICON 2015}.\hskip 1em plus 0.5em minus 0.4em\relax
  IEEE, 2015, pp. 1--5.

\bibitem[Pereira~de Figueiredo et~al.(2020)Pereira~de Figueiredo, Cardoso,
  Miranda, Moerman, Dias, and Fraidenraich]{pereira2020large}
F.~A. Pereira~de Figueiredo, F.~A. Cardoso, J.~P. Miranda, I.~Moerman, C.~F.
  Dias, and G.~Fraidenraich, ``{Large-Scale Antenna Systems and Massive Machine
  Type Communications},'' \emph{International journal of wireless information
  networks}, vol. 2020, pp. 1--23, 2020.

\bibitem[Garcia-Rodriguez et~al.(2017)Garcia-Rodriguez, Geraci, Giordano,
  Bonfante, Ding, and L{\'o}pez-P{\'e}rez]{garcia2017massive}
A.~Garcia-Rodriguez, G.~Geraci, L.~G. Giordano, A.~Bonfante, M.~Ding, and
  D.~L{\'o}pez-P{\'e}rez, ``{Massive MIMO Unlicensed: A New Approach to Dynamic
  Spectrum Access},'' \emph{IEEE Communications Magazine}, vol.~56, no.~6, pp.
  186--192, 2017.

\bibitem[{Callebaut} et~al.(2020){Callebaut}, {Gunnarsson}, {Guevara},
  {Tufvesson}, {Pollin}, {Van der Perre}, and {Johansson}]{asilomar}
G.~{Callebaut}, S.~{Gunnarsson}, A.~P. {Guevara}, F.~{Tufvesson}, S.~{Pollin},
  L.~{Van der Perre}, and A.~J. {Johansson}, ``{Massive MIMO goes Sub-GHz:
  Implementation and Experimental Exploration for LPWANs},'' in \emph{2020 54rd
  Asilomar Conference on Signals, Systems, and Computers}, 2020.

\bibitem[Ngo and Larsson(2017)]{downlink_pilots}
H.~Q. Ngo and E.~G. Larsson, ``{No Downlink Pilots Are Needed in {TDD} Massive
  {MIMO}},'' \emph{IEEE Trans Wireless Commun}, vol.~16, no.~5, pp. 2921--2935,
  May 2017.

\bibitem[{Maurer} et~al.(2007){Maurer}, {Matz}, and {Seethaler}]{4217646}
J.~{Maurer}, G.~{Matz}, and D.~{Seethaler}, ``{Low-Complexity and
  Full-Diversity MIMO Detection Based on Condition Number Thresholding},'' in
  \emph{2007 IEEE International Conference on Acoustics, Speech and Signal
  Processing - ICASSP '07}, vol.~3, 2007, pp. Iii--61--iii--64.

\bibitem[{Hoydis} et~al.(2012){Hoydis}, {Hoek}, {Wild}, and {ten
  Brink}]{6328480}
J.~{Hoydis}, C.~{Hoek}, T.~{Wild}, and S.~{ten Brink}, ``{Channel measurements
  for large antenna arrays},'' in \emph{2012 International Symposium on
  Wireless Communication Systems (ISWCS)}, 2012, pp. 811--815.

\bibitem[Bj\"{o}rnson et~al.(2017{\natexlab{b}})Bj\"{o}rnson, Hoydis, and
  Sanguinetti]{massivemimobook}
\BIBentryALTinterwordspacing
E.~Bj\"{o}rnson, J.~Hoydis, and L.~Sanguinetti, ``{Massive {MIMO} Networks:
  {Spectral}, Energy, and Hardware Efficiency},'' \emph{Foundations and
  Trends{\textregistered} in Signal Processing}, vol.~11, no. 3-4, pp.
  154--655, 2017. [Online]. Available:
  \url{http://dx.doi.org/10.1561/2000000093}
\BIBentrySTDinterwordspacing

\bibitem[Cept(2017)]{cept2017erc}
E.~Cept, ``{ERC recommendation 70-03, Relating to the use of Short Range
  Devices (SRD)},'' \emph{Electronic Communications Committee}, 2017.

\bibitem[Bj\"{o}rnson et~al.(2017{\natexlab{c}})Bj\"{o}rnson, Hoydis, and
  Sanguinetti]{bjornsonBook}
\BIBentryALTinterwordspacing
E.~Bj\"{o}rnson, J.~Hoydis, and L.~Sanguinetti, ``{Massive MIMO Networks:
  Spectral, Energy, and Hardware Efficiency},'' \emph{Found. Trends Signal
  Process.}, vol.~11, no. 3–4, p. 154–655, Nov. 2017. [Online]. Available:
  \url{https://doi.org/10.1561/2000000093}
\BIBentrySTDinterwordspacing

\bibitem[{Zhou} et~al.(2011){Zhou}, {Bai}, {Li}, {Gu}, and {Luo}]{5763372}
B.~{Zhou}, B.~{Bai}, Y.~{Li}, D.~{Gu}, and Y.~{Luo}, ``{Chordal Distance-Based
  User Selection Algorithm for the Multiuser MIMO Downlink with Perfect or
  Partial CSIT},'' in \emph{2011 IEEE International Conference on Advanced
  Information Networking and Applications}, 2011, pp. 77--82.

\bibitem[{Taniguchi} et~al.(2013){Taniguchi}, {Murata}, {Yoshida}, {Yamamoto},
  {Umehara}, {Denno}, and {Morikura}]{6831692}
M.~{Taniguchi}, H.~{Murata}, S.~{Yoshida}, K.~{Yamamoto}, D.~{Umehara},
  S.~{Denno}, and M.~{Morikura}, ``{Indoor Experiment of Multi-User MIMO User
  Selection Algorithm based on Chordal Distance},'' in \emph{2013 IEEE Global
  Communications Conference (GLOBECOM)}, 2013, pp. 3959--3964.

\bibitem[Callebaut et~al.(2019)Callebaut, Ottoy, and Van~der
  Perre]{callebaut2019cross}
G.~Callebaut, G.~Ottoy, and L.~Van~der Perre, ``{Cross-Layer Framework and
  Optimization for Efficient use of the Energy Budget of IoT Nodes},'' in
  \emph{2019 IEEE Wireless Communications and Networking Conference
  (WCNC)}.\hskip 1em plus 0.5em minus 0.4em\relax IEEE, 2019, pp. 1--6.

\end{thebibliography}

\begin{IEEEbiography}[{\includegraphics[width=1in,height=1.25in,clip,keepaspectratio]{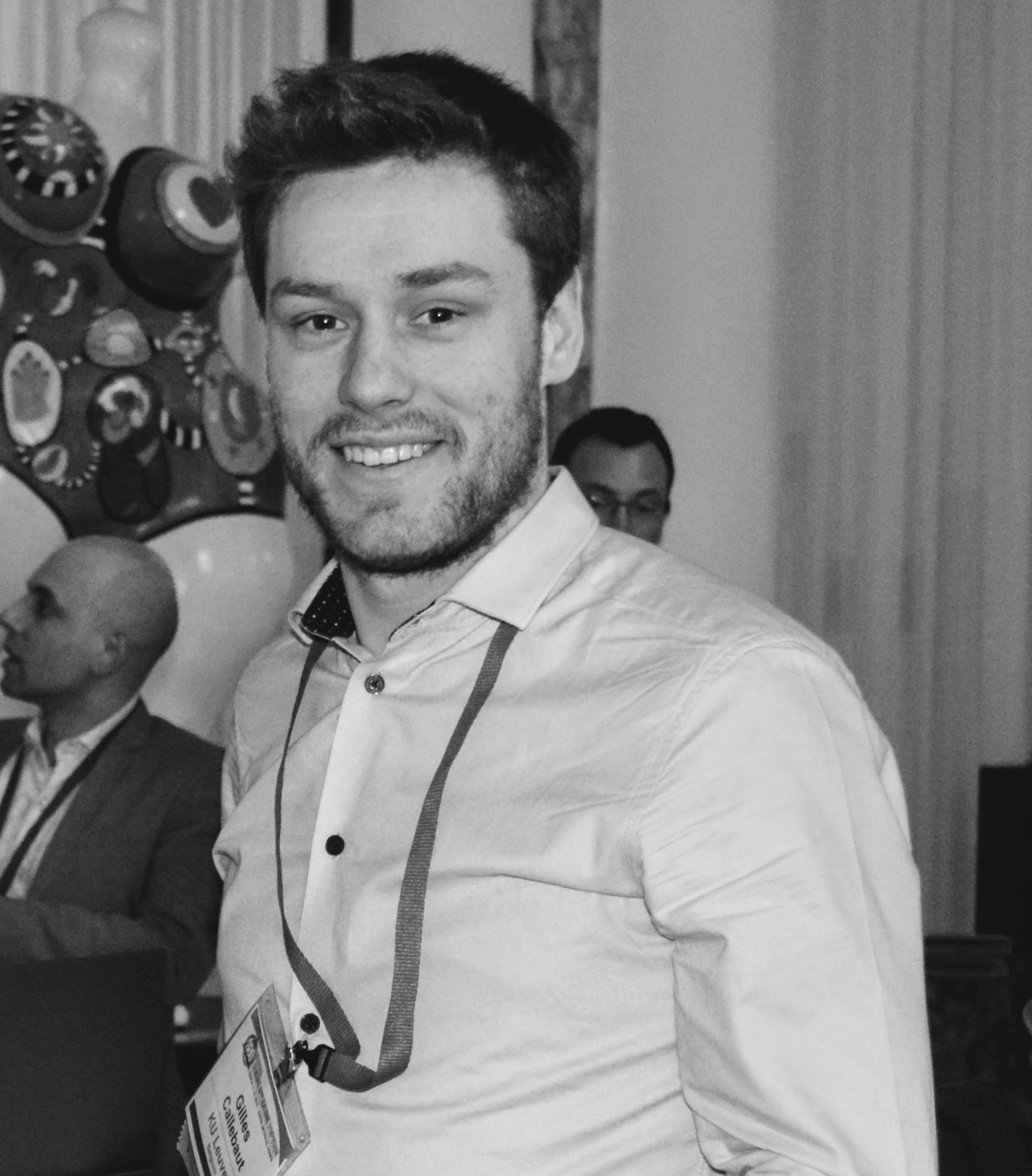}}]{\scriptsize\selectfont Gilles Callebaut}\scriptsize\selectfont
	is pursuing a Ph.D.\ in Massive MIMO for low power
	Machine Type Communication (MTC). He initiated the tutorial ``Low Power
	Wireless Technologies for Connecting Embedded Sensors in the IoT:\@ A Journey
	from Fundamentals to Hands-on''. Gilles graduated summa cum laude in 2016 and received
	the M.Sc.\ degree in engineering technology at KU Leuven campus Ghent,
	Belgium. In addition, he received the laureate award. He is currently a member
	of Dramco, a research group which is focused on wireless and mobile
	communication systems. 
\end{IEEEbiography}%
\vspace{-0.3cm}
\begin{IEEEbiography}[{\includegraphics[width=1in,height=1.25in,clip,keepaspectratio]{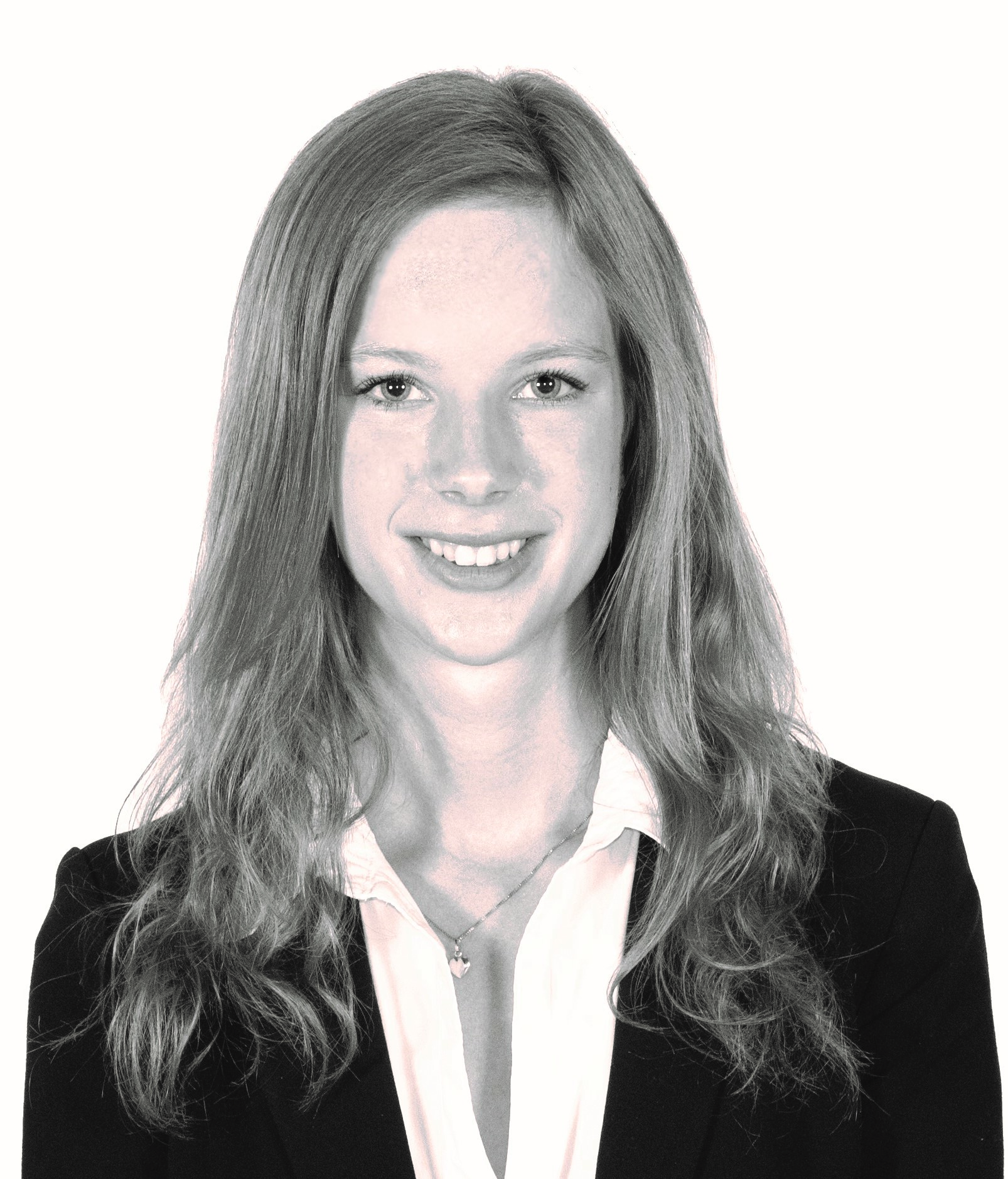}}]{\scriptsize\selectfont Sara Gunnarsson}\scriptsize\selectfont received her M.Sc. degree in Electrical Engineering from Lund University in 2017. Currently she is pursuing a double Ph.D. degree in collaboration between the Department of Electrical and Information Technology at Lund University, Sweden, and the Department of Electrical Engineering at KU Leuven, Belgium. Her main research interests are channel characterization and modeling to improve reliability and efficiency in massive MIMO systems.
\end{IEEEbiography}%
\vspace{-0.3cm}
\begin{IEEEbiography}[{\includegraphics[width=1in,height=1.25in,clip,keepaspectratio]{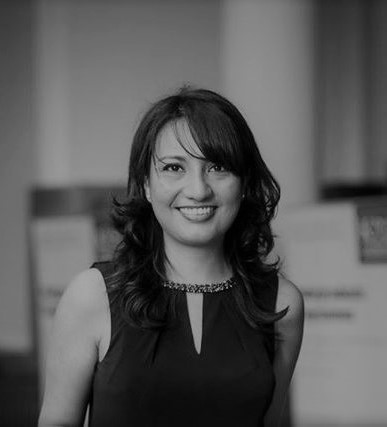}}]{\scriptsize\selectfont Andrea P. Guevara}\scriptsize\selectfont
obtained her BSc in Electronics and Telecommunications at the University of Cuenca, Ecuador in 2013. In 2015 she got her MSc in Telecommunications by Research and the prize for the best academic performance on an MSc by research programme, Faculty of Natural \& Mathematical Sciences at King's College London, UK. Since 2017 Andrea is pursuing a PhD at KU Leuven, her main interests are massive MIMO, infrastructure sharing, beamforming and power leakage.
\end{IEEEbiography}%
\vspace{-0.3cm}
\begin{IEEEbiography}[{\includegraphics[width=1in,height=1.25in,clip,keepaspectratio]{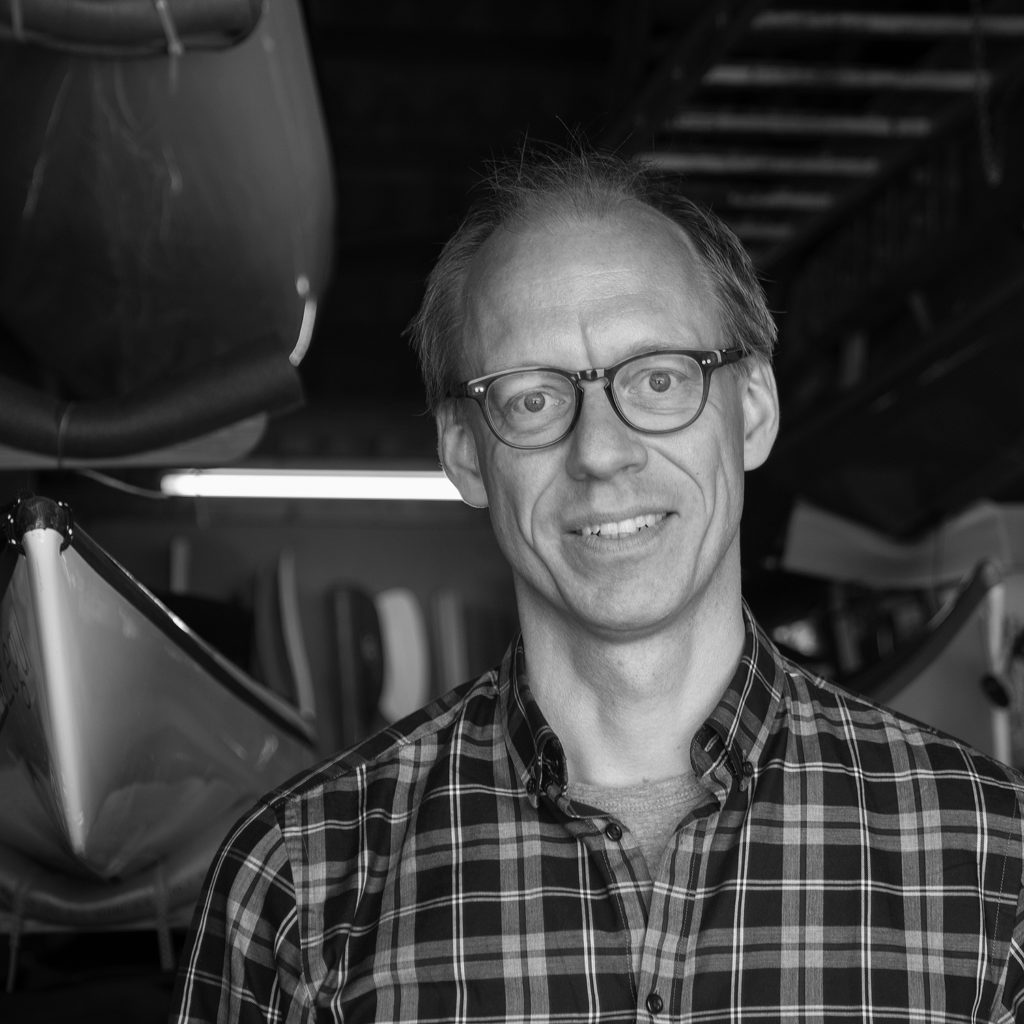}}]{\scriptsize\selectfont Anders J. Johansson}\scriptsize\selectfont
(S’98-M’05) received his
M.S., Lic. Eng. and Ph.D. degrees in electrical
engineering from Lund University, Lund, Sweden,
in 1993, 2000 and 2004 respectively.
From 1994 to 1997 he was with Ericsson Mobile
Communications AB developing transceivers and
antennas for mobile phones. Since 2005 he is an
Associate Professor at the Department of Electrical
and Information Technology, Lund University. His
research interests include antennas, wave propagation and telemetric devices for medical implants as
well as antenna systems and propagation modeling for MIMO systems.
\end{IEEEbiography}%
\vspace{-0.3cm}
\begin{IEEEbiography}[{\includegraphics[width=1in,height=1.25in,clip,keepaspectratio]{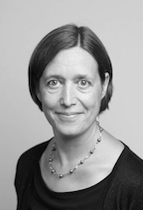}}]{\scriptsize\selectfont Liesbet Van der Perre}\scriptsize\selectfont
	is Professor at the department of Electrical Engineering at the KU Leuven in Leuven, Belgium and a guest Professor at the Electrical and Information Technology Department at Lund University,
	Sweden. Dr.\ Van der Perre was with the nano-electronics research institute
	imec in Belgium from 1997 till 2015. 
	Dr.\ Van der Perre's main research interests are in wireless communication, with a focus on physical layer and energy efficiency in both broadband communication and IoT. She was appointed honorary doctor at Lund University, Sweden, in 2015. 
\end{IEEEbiography}%
\vspace{-0.3cm}
\begin{IEEEbiography}[{\includegraphics[width=1in,height=1.25in,clip,keepaspectratio]{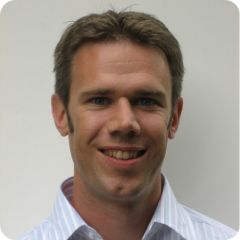}}]{\scriptsize\selectfont Fredrik Tufvesson}\scriptsize\selectfont
received the Ph.D. degree from Lund
University, Lund, Sweden, in 2000. After two years at a startup
company, he joined the Department of Electrical and Information Technology, Lund University, where he is currently
Professor of radio systems. His main research interests
include the interplay between the radio channel and the rest
of the communication system with various applications in 5G
systems. He
recently received the Neal Shepherd Memorial Award for the
best propagation paper in the IEEE Transactions on Vehicular Technology and the IEEE Communications Society Best
Tutorial Paper Award. 
\end{IEEEbiography}
\vfill
\end{document}